\documentclass[twocolumn,twoside,fleqn,showpacs,showkeys,pra,aps,10pt,superscriptaddress,final]{article}

\usepackage{hep}
\graphicspath{{figs/}}      

\def\refname{References}

\def\journalname{??}

\def\@pacs@name{PACS numbers: }%
\def\@keys@name{Keywords: }%

\def\Dated@name{Dated: }%
\def\Received@name{Received }%
\def\Revised@name{Revised }%
\def\Accepted@name{Accepted }%
\def\Published@name{Published }%
\def\address{\replace@command\address\affiliation}%
\def\altaddress{\replace@command\altaddress\altaffiliation}%

\definecolor{orangec}{cmyk}{.24,.91,.96,.18}
\definecolor{orangecc}{cmyk}{.24,.94,.96,.18}
\definecolor{oorangec}{cmyk}{.8,.2,.5,.4}
\definecolor{ooorangec}{cmyk}{1,.9,0.08,.04}      

\definecolor{orangec}{cmyk}{.15,.7,.96,.0}
\definecolor{orangecc}{cmyk}{.15,.7,.96,.0}

\newcommand{\ODDBOX}{\noindent\raisebox{-13.3cm}{\includegraphics{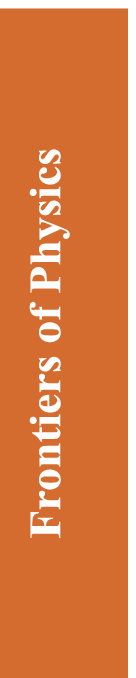}}}
\newcommand{\EVENBOX}{\noindent\raisebox{-13.3cm}{\includegraphics{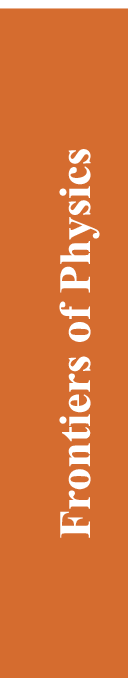}}}

\newfont{\yihao}{cmb10 at 18pt}
\newcommand{\Yihao}{\fontsize{18pt}{13.5pt}\selectfont}
\newcommand{\xiaosi}{\fontsize{11.5pt}{13.5pt}\selectfont}         

\newfont{\xbt}{cmb10 at 12pt}

\def\frontmatter@title@format{%
    \centering%
    \usefont{T1}{fradmcn}{m}{n}\yihao}%
\def\@keys@name{{\color{ooorangec}\bf Keywords~~}}%
\def\@pacs@name{{\color{ooorangec}\bf PACS numbers~~}\vspace{2mm}}%
\def\frontmatter@authorformat{\vspace{5mm}\centering\bf}%

\newcommand{\catchline}[2]{
	{\vspace*{-16.4mm}\small%
	\noindent #1\\%
	\noindent #2\\[-2mm]%
    {\color{orangec}{\rule{\textwidth}{.5pt}}}\\[2mm]
    {\color{orangec}{\Yihao\textbf{\textsc{\Papertype}}}}\\[6mm]
    }\relax\par
}

\newcommand{\doi}[1]{DOI {#1}}
\renewcommand{\title}[1]
{\vspace*{-5mm}\begin{center}
{\Yihao\bf #1}
\end{center}
}

\renewcommand{\author}[1]
{\vspace*{0mm}
\begin{center}
{\bf #1}
\end{center}
}
\newcommand{\add}[1]{\begin{center}{\small\it #1}\end{center}}

\newcommand{\abs}[1]{
\begin{center}
\parbox[t]{156mm}{\noindent\color{oorangec}#1}
\end{center}}

\newcommand{\keywords}[1]{
\begin{center}
\parbox[t]{156mm}{\noindent{\bf\color{ooorangec}Keywords}\ \ #1}
\end{center}}

\newcommand{\pacsnumbers}[1]{
\begin{center}
\parbox[t]{156mm}{\noindent{\bf\color{ooorangec}PACS numbers}\ \ #1\vspace*{5mm}}
\end{center}}

\newcommand{\acknowledgements}[1]{\vspace*{4mm}\noindent{\renewcommand{\baselinestretch}{1.05}\footnotesize{\color{ooorangec}\bf Acknowledgements}\quad{#1}}}

\def\journalname{??}
\def\volumenumber#1{\gdef\@volumenumber{#1}}%
\def\@volumenumber{}%
\def\issuenumber#1{\gdef\@issuenumber{#1}}%
\def\@issuenumber{}%
\def\volumeyear#1{\gdef\@volumeyear{#1}}%
\def\@volumeyear{}%

\renewcommand\thesection{\arabic{section}}
\renewcommand\thesubsection{\arabic{section}.\arabic{subsection}}
\renewcommand\thesubsubsection{\arabic{section}.\arabic{subsection}.\arabic{subsubsection}}

\titleformat{\section}[hang]{\color{ooorangec}\vspace*{-1.2mm}\titlerule\vspace{1mm}\large\usefont{T1}{fradmcn}{m}{n}\xbt}{\thesection}{1em}{}
\titlespacing{\section}{0mm}{8mm}{5mm}

\titleformat{\subsection}{\normalfont\normalsize\color{ooorangec}}{\thesubsection}{1em}{}
\titlespacing{\subsection}{0mm}{5mm}{3mm}

\titleformat{\subsubsection}{\normalfont\normalsize\it\color{ooorangec}}{\thesubsubsection}{1em}{}
\titlespacing{\subsubsection}{0mm}{3mm}{3mm}

\titlecontents{section}
[0em] {\normalfont} {\thecontentslabel\quad} {\thecontentslabel}
{\titlerule*[.7pc]{}\contentspage}

\titlecontents{subsection}
[1.5em] {\normalfont} {\thecontentslabel\quad} {\thecontentslabel}
{\titlerule*[.7pc]{}\contentspage}

\titlecontents{subsubsection}
[4em] {\normalfont} {\thecontentslabel\quad} {\thecontentslabel}
{\titlerule*[.7pc]{}\contentspage}

\usepackage[small]{caption2}

\newlength{\halfpagewidth}
\divide\halfpagewidth by 1

\usepackage{hyperref}
\usepackage{mathrsfs}
\usepackage{ulem}
\usepackage[figuresright]{rotating}

\def\prd{Phys. Rev. D}
\def\prl{Phys. Rev. Lett.}
\def\aap{Astron. Astroph.}
\def\apj{Astrophys. J.}
\def\apjl{Astrophys. J. Lett.}
\def\apjs{Astrophys. J. Suppl.}
\def\mnras{Mon. Not. R. Astron. Soc.}
\def\nar{New Astron. Rev.}
\def\nat{Nature}
\def\jcap{JCAP}

\begin{document}

\newcommand{\Papertype}{\sc Review article} 
\def\volumeyear{2021} 
\def\volumenumber{16(4)} 
\def\issuenumber{44300} 
\def\journalname{Front. Phys.} 
\newcommand{\doiurl}{https://doi.org/10.1007/s11467-020-1049-x} 
\newcommand{\allauthors}{Jun-Jie Wei and Xue-Feng Wu} 

\twocolumn[
\begin{@twocolumnfalse}
\catchline{\journalname~\volumenumber,~\issuenumber~(\volumeyear)}{\doi{\doiurl}} 
\thispagestyle{firstpage}


\title{Testing Fundamental Physics with Astrophysical Transients}

\author{Jun-Jie Wei$^{1,2,*}$, Xue-Feng Wu$^{1,2,\dag}$}

\add{1. Purple Mountain Observatory, Chinese Academy of Sciences, Nanjing 210023, China\\
2. School of Astronomy and Space Sciences, University of Science and Technology of China, Hefei 230026, China\\
Corresponding authors.\ E-mail: $^*$jjwei@pmo.ac.cn, $^\dag$xfwu@pmo.ac.cn\\}

\abs{Explosive astrophysical transients at cosmological distances can be used to place precision tests of
the basic assumptions of relativity theory, such as Lorentz invariance, the photon zero-mass hypothesis,
and the weak equivalence principle (WEP). Signatures of Lorentz invariance violations (LIV) include vacuum
dispersion and vacuum birefringence. Sensitive searches for LIV using astrophysical sources such as
gamma-ray bursts, active galactic nuclei, and pulsars are discussed. The most direct consequence of a
nonzero photon rest mass is a frequency dependence in the velocity of light propagating in vacuum.
A detailed representation of how to obtain a combined severe limit on the photon mass using fast radio bursts
at different redshifts through the dispersion method is presented. The accuracy of the WEP has been well
tested based on the Shapiro time delay of astrophysical messengers traveling through a gravitational field.
Some caveats of Shapiro delay tests are discussed. In this article, we review and update the status of
astrophysical tests of fundamental physics.}


\pacsnumbers{04.60.Pp, 95.30.Sf, 98.70.Dk, 98.70.Rz}

\vspace*{-6mm}

\end{@twocolumnfalse}
]


\tableofcontents

\section{
Introduction} \label{sec:intro}
Einstein's theory of special and general relativity is a major pillar of modern physics, with a wide application
in astrophysics. It is, therefore, of great scientific significance to test the validity of the basic assumptions
of relativity theory, such as Lorentz invariance, the photon zero-mass hypothesis, and the weak equivalence principle (WEP).
When the terrestrial conditions eventually impose some limitations, the extreme features of astrophysical phenomena
afford the ideal testbeds for obtaining higher precision tests of the fundamental laws of physics.

Lorentz invariance, which is the fundamental symmetry of Einstein's special relativity, says that
the relevant physical laws of a non-accelerating physical system are invariant under Lorentz transformation.
However, deviations from Lorentz invariance at a natural energy scale are suggested in many quantum gravity
(QG) theories seeking to unify general relativity and quantum mechanics, such as loop quantum gravity
\cite{1999PhRvD..59l4021G,2002PhRvD..65j3509A}, double special relativity
\cite{2002IJMPD..11...35A,2002Natur.418...34A,2002PhLB..539..126K,2003PhRvD..67d4017M}, and superstring
theory \cite{1989PhRvD..39..683K}. This natural scale, referred to as the ``QG energy scale'' $E_{\rm QG}$,
is generally supposed to be around the Planck energy $E_{\rm Pl}=\sqrt{\hbar c^{5}/G}\simeq1.22\times10^{19}$
GeV \cite{1989PhRvD..39..683K,1991NuPhB.359..545K,1995PhRvD..51.3923K,
2005LRR.....8....5M,2005hep.ph....6054B,2013LRR....16....5A,2014RPPh...77f2901T}.
The dedicated experimental tests of Lorentz invariance may therefore help to clear the path to
a grand unified theory. A compilation of various recent experimental tests may be found in
Ref. \cite{2011RvMP...83...11K}. Although any violations of Lorentz symmetry are predicted to
be tiny at observable energies $\ll E_{\rm Pl}$, they can accumulate to measurable levels
over large distances. Therefore, astrophysical measurements involving long baselines can
provide sensitive probes of Lorentz invariance violation (LIV). In the photon sector,
the potential signatures of LIV include vacuum dispersion and vacuum birefringence \cite{2008ApJ...689L...1K}.
Vacuum dispersion would produce an energy-dependent speed of light that in turn would translate
into differences in the arrival time of photons with different energies traveling over cosmological
distances. Lorentz invariance can therefore be tested using astrophysical time-of-flight measurements
(e.g., Refs.
\cite{1998Natur.393..763A,
2005PhLB..625...13P,
2006APh....25..402E,
2008JCAP...01..031J,
2008ApJ...689L...1K,
2009PhRvD..80a5020K,
2009Sci...323.1688A,
2009Natur.462..331A,
2012APh....36...47C,
2012PhRvL.108w1103N,
2013PhRvD..87l2001V,
2013APh....43...50E,
2015PhRvD..92d5016K,
2015APh....61..108Z,
2017ApJ...834L..13W,
2017ApJ...842..115W,
2017ApJ...851..127W,
2019PhRvD..99h3009E,
PhysRevLett.125.021301,
1999PhRvL..83.2108B,
Kaaret1999}).
Similarly, the effect of vacuum birefringence would accumulate over astrophysical distances resulting in
a detectable rotation of the polarization vector of linearly polarized emission as a function of energy.
Thus, Lorentz invariance can also be probed with astrophysical polarization measurements (e.g., Refs.
\cite{1990PhRvD..41.1231C,
1998PhRvD..58k6002C,
2001PhRvD..64h3007G,
2001PhRvL..87y1304K,
2006PhRvL..97n0401K,
2007PhRvL..99a1601K,
2013PhRvL.110t1601K,
2003Natur.426Q.139M,
2004PhRvL..93b1101J,
2007MNRAS.376.1857F,
2009JCAP...08..021G,
2011PhRvD..83l1301L,
2011APh....35...95S,
2012PhRvL.109x1104T,
2013MNRAS.431.3550G,
2014MNRAS.444.2776G,
2016MNRAS.463..375L,
2017PhRvD..95h3013K,
2019PhRvD..99c5045F,
2019MNRAS.485.2401W}).
Typically, polarization measurements place more stringent constraints on LIV. This can be understood by
the fact that polarization measurements are more sensitive than time-of-flight measurements by a factor
$\propto1/E$, where $E$ is the energy of the photon \cite{2009PhRvD..80a5020K}. However, numerous predicted
Lorentz-violating signals have no vacuum birefringence, so constraints from time-of-flight measurements
are indispensable in an extensive search for nonbirefringent effects.

The postulate that all electromagnetic waves travel in vacuum at the constant speed $c$ is one of the
foundations of Maxwell's electromagnetism and Einstein's special relativity. The constancy of light speed
implies that the quantum of light, or photon, should be massless. The validity of this postulate can
therefore be tested by searching for a rest mass of the photon. However, none of the experiments so far could
confirm that the photon rest mass is absolutely zero. Based on the uncertainty principle, when using
the age of the universe ($T\sim10^{10}$ yr), there is an ultimate upper limit on the photon rest mass,
i.e., $m_{\gamma}\leq\hbar/Tc^2\approx10^{-69}~\rm{kg}$ \cite{1971RvMP...43..277G,2005RPPh...68...77T}.
The best one can hope to do is to set ever tighter limits on $m_{\gamma}$ and push the experimental results
even closer to the ultimate upper limit. In theory, a nonzero photon rest mass can be accommodated into
electromagnetism through the Proca equations. Using them, some possible visible effects associated with
a massive photon have been carefully studied, which open the door to useful approaches for terrestrial
experiments or astrophysical observations aimed at placing upper limits on the photon rest mass (see Refs.
\cite{1971RvMP...43..277G,2005RPPh...68...77T,2006AcPPB..37..565O,2010RvMP...82..939G,2011EPJD...61..531S}
for reviews). To date, the experimental methods for constraining $m_{\gamma}$ include
the frequency dependence of the speed of light ($m_{\gamma}\leq5.1\times10^{-51}~\mathrm{kg}$)
\cite{1964Natur.202..377L,1969Natur.222..157W,1999PhRvL..82.4964S,2016JHEAp..11...20Z,2017RAA....17...13W,
2016ApJ...822L..15W,2016PhLB..757..548B,2017PhLB..768..326B,2017PhRvD..95l3010S,2018JCAP...07..045W,
2019ApJ...882L..13X,2020arXiv200609680W}, Coulomb's inverse square law ($m_{\gamma}\leq1.6\times10^{-50}~\mathrm{kg}$)
\cite{1971PhRvL..26..721W}, Amp$\rm \grave{e}$re's law ($m_{\gamma}\leq(8.4\pm0.8)\times10^{-49}~\mathrm{kg}$)
\cite{1992PhRvL..68.3383C}, torsion balance ($m_{\gamma}\leq1.2\times10^{-54}~\mathrm{kg}$)
\cite{1998PhRvL..80.1826L,2003PhRvL..91n9101G,2003PhRvL..90h1801L,2003PhRvL..91n9102L},
gravitational deflection of electromagnetic waves ($m_{\gamma}\leq10^{-43}~\mathrm{kg}$)
\cite{1973PhRvD...8.2349L,2004PhRvD..69j7501A}, Jupiter's magnetic field ($m_{\gamma}\leq8\times10^{-52}~\mathrm{kg}$)
\cite{1975PhRvL..35.1402D}, magnetohydrodynamic phenomena of the solar wind
($m_{\gamma}\leq1.4\times10^{-49}-3.4\times10^{-51}~\mathrm{kg}$) \cite{1997PPCF...39...73R,2007PPCF...49..429R,2016APh....82...49R},
cosmic magnetic fields ($m_{\gamma}\leq10^{-62}~\mathrm{kg}$) \cite{1959PThPS..11....1Y,1976UsFiN.119..551C,PhysRevLett.98.010402},
supermassive black-hole spin ($m_{\gamma}\leq10^{-56}-10^{-58}~\mathrm{kg}$) \cite{2012PhRvL.109m1102P},
spindown of a white-dwarf pulsar ($m_{\gamma}\leq(6.3-9.6)\times10^{-53}~\mathrm{kg}$) \cite{2017ApJ...842...23Y},
and so on. Among these methods, the most direct and robust one is to detect a possible frequency dependence
in the velocity of light. In this article, we will review the photon mass limits from the dispersion of
electromagnetic waves of astrophysical sources.

Einstein's WEP states that any freely falling, uncharged test body will follow a trajectory,
independent of its internal structure and composition \cite{2006LRR.....9....3W,2014LRR....17....4W}.
It is the basic ingredient of general relativity and other metric theories of gravity. The most famous
tests of the WEP are the E$\ddot{\rm o}$tv$\ddot{\rm o}$s-type experiments, which compare the accelerations
of two laboratory-sized objects consisted of different composition in a known gravitational field (see
Ref. \cite{2014LRR....17....4W} for a review). For laboratory-sized objects with macroscale masses,
the accuracy of the WEP can be tested in a Newtonian context. However, the motion of test particles
(like photons or neutrinos) in a gravitational field is not precisely described by Newtonian dynamics.
As such, the parameterized post-Newtonian (PPN) formalism has been developed to describe exactly
their motion. Each theory of gravitation satisfying the WEP is specified by a set of PPN parameters.
The WEP can then also be tested by massless (or negligible rest-mass) particles in the context of
the PPN formalism and any possible deviation from WEP is characterized by the PPN parameter, $\gamma$,
of a particular gravity theory. Here $\gamma$ reflects the level of space curved by unit rest mass
\cite{2006LRR.....9....3W,2014LRR....17....4W}. In general relativity, $\gamma$ is predicted to be strictly 1.
The determination of the absolute $\gamma$ value has reached high precision. The light deflection
measurements through very-long-baseline radio interferometry yielded an agreement with general relativity
to 0.01 percent, i.e., $\gamma-1=(-0.8\pm1.2)\times10^{-4}$ \cite{2009A&A...499..331L,2011A&A...529A..70L}.
The radar time-delay measurement from the Cassini spacecraft obtained a more stringent constraint
$\gamma-1=(2.1\pm2.3)\times10^{-5}$ \cite{2003Natur.425..374B}. Regardless of the absolute value
of $\gamma$, all metric theories of gravity incorporating the WEP predict that different species of
messenger particles (photons, neutrinos, and gravitational waves (GWs)), or the same species of particles
but with different internal properties (e.g, energies or polarization states), traveling through
the same gravitational fields, must follow identical trajectories and undergo the same $\gamma$-dependent
Shapiro delay \cite{1964PhRvL..13..789S}. To test the WEP, therefore, the issue is not whether the value
of $\gamma$ is very nearly unity, but whether it is the same for all test particles. The Shapiro time delay
of test particles emitted from the same astrophysical sources has been widely applied to constrain
a possible violation of the WEP through the relative differential variations of the $\gamma$ values (e.g.,
\cite{1988PhRvL..60..173L,1988PhRvL..60..176K,2015ApJ...810..121G,2015PhRvL.115z1101W,2016PhRvD..94b4061W,2017MNRAS.469L..36Y,2019PhRvD..99j3012W}).

In this review, we summarize the current status on astrophysical tests of fundamental physics
and attempt to chart the future of the subject. In Section~\ref{sec:LIV}, we review two common astrophysical
approaches to testing the LIV effects. Section~\ref{sec:photonmass} is dedicated to discuss astrophysical bounds on
the photon rest mass through the measurement of the frequency dependence of the speed of light.
In Section~\ref{sec:WEP}, we focus on tests of the WEP through the Shapiro (gravitational) time delay effect.
Finally, a summary and future prospect are presented in Section~\ref{sec:summary}.

\section{
Astrophysical tests of LIV}\label{sec:LIV}

Various QG theories that extend beyond the standard model predict violations of Lorentz invariance at energies
approaching the Planck scale \cite{1989PhRvD..39..683K,1991NuPhB.359..545K,1995PhRvD..51.3923K,
2005LRR.....8....5M,2005hep.ph....6054B,2013LRR....16....5A,2014RPPh...77f2901T}. The existence of LIV can produces
an energy-dependent vacuum dispersion of light, which leads to time delays between promptly emitted photons with
different energies, as well as an energy-dependent rotation of the polarization plane of linearly polarized photons
resulting from vacuum birefringence. In this section, we first review the recent achievements in sensitivity of
vacuum dispersion time-of-flight measurements. Then the progress on LIV limits from astrophysical polarization
measurements is discussed.

\subsection{Vacuum dispersion from LIV}\label{subsec:LIV1}
\subsubsection{General formulae}
Several QG theories proposed to describe quantum spacetime, predict the granularity of space and time.
For example, string theory that remove the point-like nature of the particles by introducing to each of them
a (mono)-dimensional extension: the string (see, e.g., Smolin \cite{2003hep.th....3185S} for reviews).
Loop quantum gravity that question the continuousness and smoothness of spacetime quantizing it into discrete
energy levels like those observed in the classical quantum-mechanical systems to construct a complex geometric
structure called spin networks (see, e.g., Rovelli \cite{1998LRR.....1....1R} for reviews).
In both proposed theories emerge a minimal length for physical space (and time), although with different and
somewhat opposite theoretical approaches \cite{2019arXiv191102154B}. The minimal spatial and temporal scales
associated to the quantization are in terms of standard units: $\textit{l}_{\rm P}\sim\sqrt{\hbar G/c^3}\sim10^{-33}$ cm
and $t_{\rm P}\sim\sqrt{\hbar G/c^5}\sim10^{-43}$ s for the Planck length and time, respectively.
A significant class of QG theories predict that the propagation of photons through this discrete spacetime
might exhibit a non-trivial dispersion relation in vacuum \cite{1998Natur.393..763A}, corresponding to
the possible violation or break of the Lorentz invariance via an energy-dependent speed of light.
In vacuum, the LIV-induced modifications to the dispersion relation of photons can be described using
a Taylor series expansion:
\begin{equation}
E^{2}\simeq p^{2}c^{2}\left[1-\sum_{n=1}^{\infty}s_{\pm}\left(\frac{E}{E_{{\rm QG},n}}\right)^{n}\right]\;,
\label{eq:LIVdispersion}
\end{equation}
where $p$ is the photon momentum, $E_{\rm QG}$ is the hypothetical energy scale at which QG effects would become significant,
and $s_{\pm}=\pm1$ is a theory-dependent factor. For $E\ll E_{\rm QG}$, the sum is dominated by the lowest-order term
in the series. Considering only the lowest-order dominant term, the photon group velocity can therefore be expressed by
\begin{equation}
v(E)=\frac{\partial E}{\partial p}\approx c\left[1-s_{\pm}\frac{n+1}{2}\left(\frac{E}{E_{{\rm QG},n}}\right)^{n}\right]\;,
\end{equation}
where $n=1$ and $n=2$ correspond to the linear and quadratic LIV, respectively. The coefficient $s_{\pm}=+1$ ($s_{\pm}=-1$)
stands for a decrease (increase) in the photon speed along with increasing photon energy (also refers to the ``subluminal''
and ``superluminal'' scenarios).

Because of the energy dependence of $v(E)$, two photons with different observer-frame energies ($E_{h}>E_{l}$)
emitted simultaneously from the same astrophysical source at redshift $z$ would arrive at the observer at different times.
Taking account of the cosmological expansion, we derive the expression of the LIV-induced time delay \cite{2008JCAP...01..031J}:
\begin{equation}
\begin{aligned}
\Delta t_{\rm LIV}&=t_{h}-t_{l}\\
                  &=s_{\pm}\frac{1+n}{2H_{0}}\frac{E_{h}^{n}-E_{l}^{n}}{E_{{\rm QG}, n}^{n}}
\int_{0}^{z}\frac{(1+z')^{n}dz'}{\sqrt{\Omega_{\rm m}(1+z')^{3}+\Omega_{\Lambda}}}\;,
\label{eq:tLIV}
\end{aligned}
\end{equation}
where $t_{h}$ and $t_{l}$, respectively, denote the arrival times of the high-energy and low-energy photons,
$H_{0}$ is the Hubble constant, and $\Omega_{\rm m}$ and $\Omega_{\Lambda}$ are the matter energy density and
the vacuum energy density (parameters of the flat $\Lambda$CDM model).
Since the adopted energy range generally spans several orders of magnitude, one can approximate
$(E_{h}^{n}-E_{l}^{n})\approx E_{h}^{n}$. Adopting $z=20$ as a firm upper limit for the redshift of any source,
we find the LIV-induced time delay is $|\Delta t_{\rm LIV}|\leq 0.5(E_{h\;{\rm GeV}}/\zeta)$ s for the linear
($n=1$) LIV case and $|\Delta t_{\rm LIV}|\leq 5.2\times10^{-19}(E_{h\;{\rm GeV}}/\zeta)^{2}$ s for the quadratic
($n=2$) LIV case, where $E_{h\;{\rm GeV}}=E_{h}/$(1 GeV) and $\zeta=E_{\rm QG}/E_{\rm Pl}$. These indicate that
first order ($n=1$) effects would lead to potentially observable delays, while second order ($n=2$) effects
are so tiny that it would be impossible to observe them with this time-of-flight technique.

\subsubsection{Present results}
Eq. (\ref{eq:tLIV}) shows that the LIV-induced time delay increases with the energy of the photons and the distance of
the source. A short timescale of the signal variability easily provides a reference time to measure the time delay.
The higher the photon energy, the longer the source distance, and the shorter the time delay, the more stringent limits
on the QG energy scale one can reach. Exploiting the rapid variations of gamma-ray emissions from astrophysical
sources at large distances to constrain LIV was first proposed by Amelino-Camelia et al. \cite{1998Natur.393..763A}.
At present, tests of LIV have been made using the observations of gamma-ray bursts (GRBs), active galactic nuclei (AGNs),
and pulsars.


\begin{itemize}
  \item Gamma-ray bursts

\end{itemize}

\begin{figure*}
\centering
\includegraphics[width=0.45\textwidth]{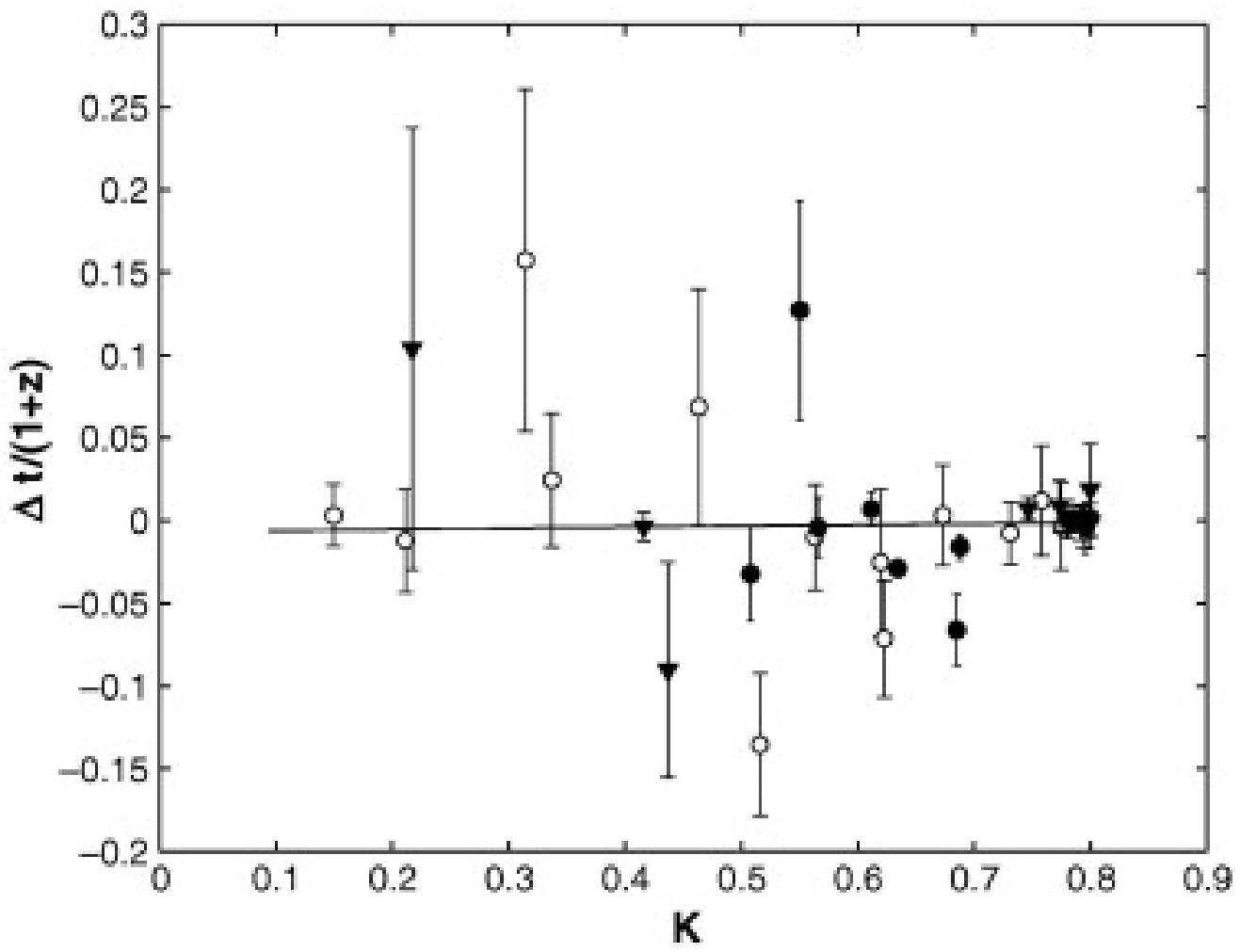}
\includegraphics[width=0.45\textwidth]{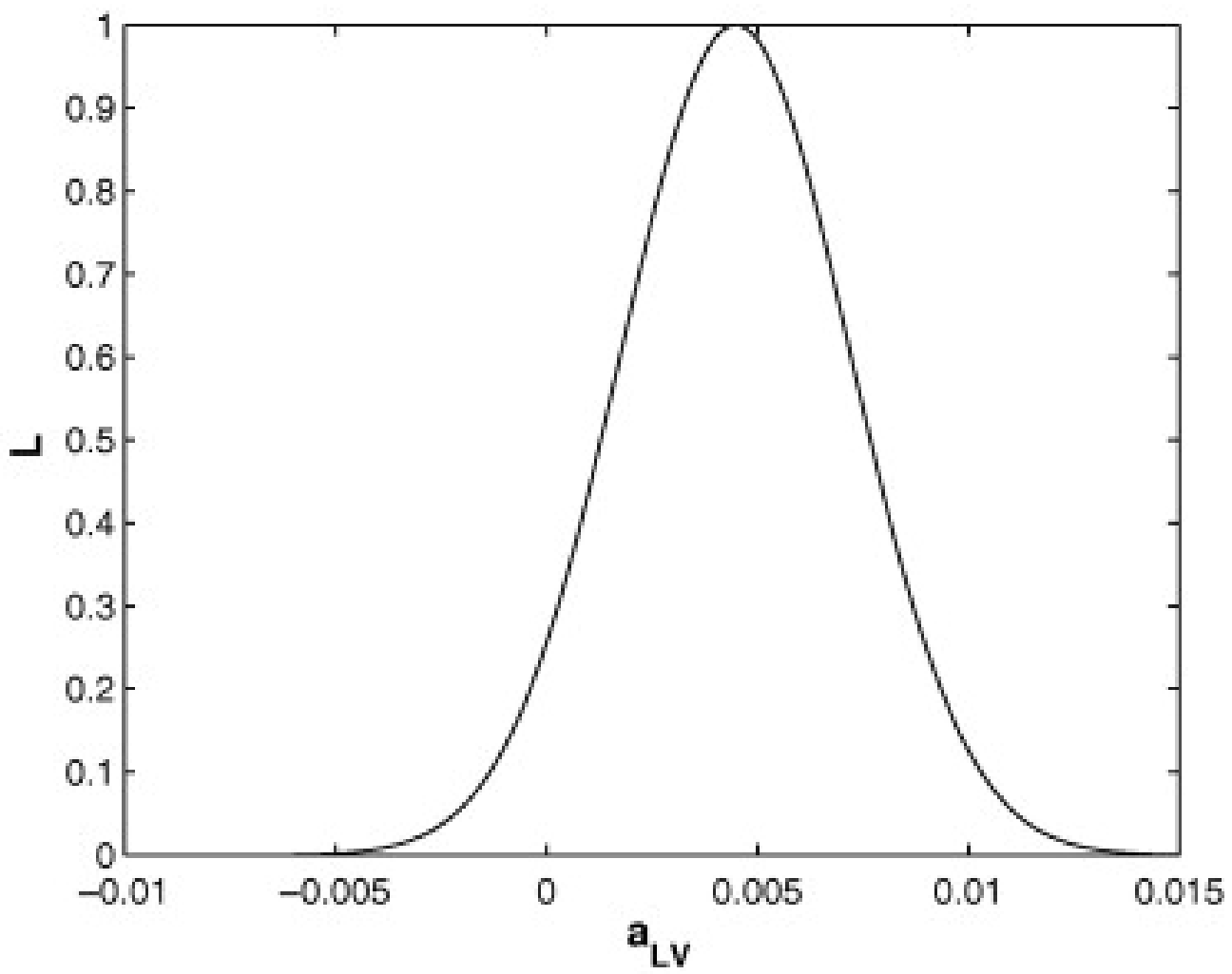}
\caption{Left panel: the rescaled time lags between two selected observer-frame energy bands from the full set
of 35 GRBs with known redshifts observed until 2006 by BATSE (closed circles), HETE-2 (open circles) and Swift (triangles).
Right panel: the likelihood function for the slope parameter $a_{\rm LV}$. Reproduced from Ref. \cite{2008APh....29..158E}.}
\label{fig:LIV1}
\end{figure*}

GRBs are among the most distant gamma-ray sources and their signals vary on subsecond timescales.
As such, GRBs are identified as promising sources for LIV studies.
There have been many searches for LIV through studies on the time-of-flight measurements of GRBs. Some of the pre-Fermi
studies are those by Ellis et al. \cite{Ellis2003} using BATSE GRBs;
by Boggs et al. \cite{2004ApJ...611L..77B} using RHESSI observations of GRB 021206; by Ellis et al.
\cite{2006APh....25..402E,2008APh....29..158E} using BATSE, HETE-2, and Swift GRBs; by Rodr{\'\i}guez-Mart{\'\i}nez et al.
\cite{2006JCAP...05..017R} using Konus-Wind and Swift observations of GRB 051221A; by Bolmont et al.
\cite{2008ApJ...676..532B} using HETE-2 GRBs; and by Lamon et al. \cite{2008GReGr..40.1731L} using INTEGRAL GRBs.
Because of the unprecedented sensitivity for detecting the prompt high-energy GRB emission (up to tens of GeV)
by the Fermi Large Area Telescope (LAT), more stringent constraints on LIV have been obtained using Fermi observations.
These constraints include those by the Fermi Collaboration using GRBs 080916C \cite{2009Sci...323.1688A} and
090510 \cite{2009Natur.462..331A}; by Xiao \& Ma \cite{2009PhRvD..80k6005X} using GRB 090510; and by Refs.
\cite{2010APh....33..312S,2012APh....36...47C,2012PhRvL.108w1103N,2013PhRvD..87l2001V,2015APh....61..108Z,
2016APh....82...72X,2016PhLB..760..602X,2018JCAP...01..050X,Liu2018,2019PhRvD..99h3009E}
using multiple Fermi GRBs. Particularly, Abdo et al. \cite{2009Natur.462..331A} used the highest energy (31 GeV)
photon of the short GRB 090510 detected by Fermi/LAT to constrain the linear LIV energy scale ($E_{{\rm QG}, 1}$).
The burst has a redshift $z=0.903$. This 31 GeV photon was detected 0.829 s after the Fermi Gamma-Ray Burst Monitor
(GBM) trigger. If the 31 GeV photon was emitted at the beginning of the first GBM pulse, one has $\Delta t_{\rm LIV}<859$
ms, which gives the most conservative constraint $E_{{\rm QG}, 1}>1.19E_{\rm Pl}$. If the 31 GeV photon is associated
with the contemporaneous $<1$ MeV spike, one has $\Delta t_{\rm LIV}<10$ ms, which gives the least conservative constraint
$E_{{\rm QG}, 1}>102E_{\rm Pl}$. We can see that the linear LIV models requiring $E_{{\rm QG}, 1}\leq E_{\rm Pl}$ are
disfavored by these results. Subsequently, Vasileiou et al. \cite{2013PhRvD..87l2001V} used three statistical techniques
to constrain the total degree of dispersion in the data of four LAT-detected GRBs. For the subluminal case, their most
stringent limits are derived from GRB 090510 and are $E_{{\rm QG}, 1}>7.6E_{\rm Pl}$ and
$E_{{\rm QG}, 2}>1.3\times10^{11}~\mathrm{GeV}$ for linear and quadratic LIV, respectively. These limits improve previous
constraints by a factor of $\sim2$.
Recently, the Major Atmospheric Gamma Imaging Cherenkov (MAGIC) telescopes first detected GRB 190114C in the sub-TeV
energy domain (i.e., 0.2---1 TeV), recording the highest energy photons ever observed from a GRB \cite{2019Natur.575..455M}.
Using conservative assumptions on the possible intrinsic spectral and temporal emission properties, the MAGIC
Collaboration searched for an energy dependence in the arrival time of the most energetic photons and presented
competitive limits on the quadratic leading-order LIV-induced vacuum dispersion \cite{PhysRevLett.125.021301}.
The resulting constraints from GRB 190114C are $E_{{\rm QG}, 2}>6.3\times10^{10}~\mathrm{GeV}$
and $E_{{\rm QG}, 2}>5.6\times10^{10}~\mathrm{GeV}$ for the subluminal and superluminal cases, respectively.

By using the arrival-time differences between high-energy and low-energy photons (the so-called spectral lags) from GRBs,
Lorentz invariance has been tested with unprecedented accuracy. Such time-of-flight tests, however, are subject to a bias
related to a possible intrinsic time lag introduced from the unknown emission mechanism of the sources, which would enhance
or cancel-out the delay caused by the LIV effects. That is, the method of the arrival-time difference is tempered by our
ignorance concerning potential source-intrinsic effects. The first attempt to mitigate the intrinsic time lag problem was
proposed by Ellis et al. \cite{Ellis2003,2006APh....25..402E,2008APh....29..158E}, who suggested working on a large sample
of GRBs with different redshifts. For each GRB, Ellis et al. \cite{2006APh....25..402E} looked for the spectral lag
in the light curves recorded in the chosen observer-frame energy bands 115--320 and 25--55 keV.
To account for the poorly known intrinsic time lag, they fitted the observed spectral lags of 35 GRBs with the inclusion of
a constant $b_{\rm sf}$ specified in the rest frame of the source. The observed arrival time delays thus have two
contributions $\Delta t_{\rm obs}=\Delta t_{\rm LIV}+b_{\rm sf}(1+z)$, reflecting the possible LIV effects and
source-intrinsic effects \cite{2006APh....25..402E}. Rescaling $\Delta t_{\rm obs}$ by a factor $(1+z)$, then one has
a simple linear fitting function
\begin{equation}
\frac{\Delta t_{\rm obs}}{1+z}=a_{\rm LV}K+b_{\rm sf}\;,
\label{eq:LIV-linear}
\end{equation}
where
\begin{equation}
K=\frac{1}{(1+z)}\int_{0}^{z}\frac{(1+z')dz'}{h(z')}
\end{equation}
is a function of redshift which depends on the cosmological model,
the slope $a_{\rm LV}=\Delta E/(H_0 E_{\rm QG})$ is related to the QG energy scale, and
the intercept $b_{\rm sf}$ denotes the possible unknown intrinsic time lag.
In the standard flat $\Lambda$CDM model, the dimensionless expansion rate $h(z)$ is expressed as
$h(z)=\sqrt{\Omega_{\rm m}(1+z)^{3}+\Omega_{\Lambda}}$. Note that the linear LIV ($n=1$) in the
subluminal case ($s_{\pm}=+1$) was considered in this work. Within a frame work of the concordance $\Lambda$CDM model,
a linear fit to the rescaled time lags extracted from 35 light curve pairs is shown in the left panel of
Figure~\ref{fig:LIV1}. The best-fit line corresponds to $\frac{\Delta t_{\rm obs}}{1+z}=(0.0068\pm0.0067)K-(0.0065\pm0.0046)$
and the likelihood function for the slope parameter $a_{\rm LV}$ is presented in the right panel
of Figure~\ref{fig:LIV1}. The 95\% confidence-level lower limit on the linear LIV energy scale derived
from the likelihood function of $a_{\rm LV}$ is $E_{\rm QG}\geq1.4\times10^{16}$ GeV \cite{2008APh....29..158E}.
Going beyond the $\Lambda$CDM cosmology, Refs. \cite{2009CQGra..26l5007B,2015ApJ...808...78P} extended this analysis
to different cosmological models and showed that the result is insensitive to the adopted background cosmology.
Subsequently, some cosmology-independent approaches were applied to probe the possible LIV effects \cite{2018PhLB..776..284Z,2020ApJ...890..169P}.

\begin{figure*}
\centering
\includegraphics[width=0.8\textwidth]{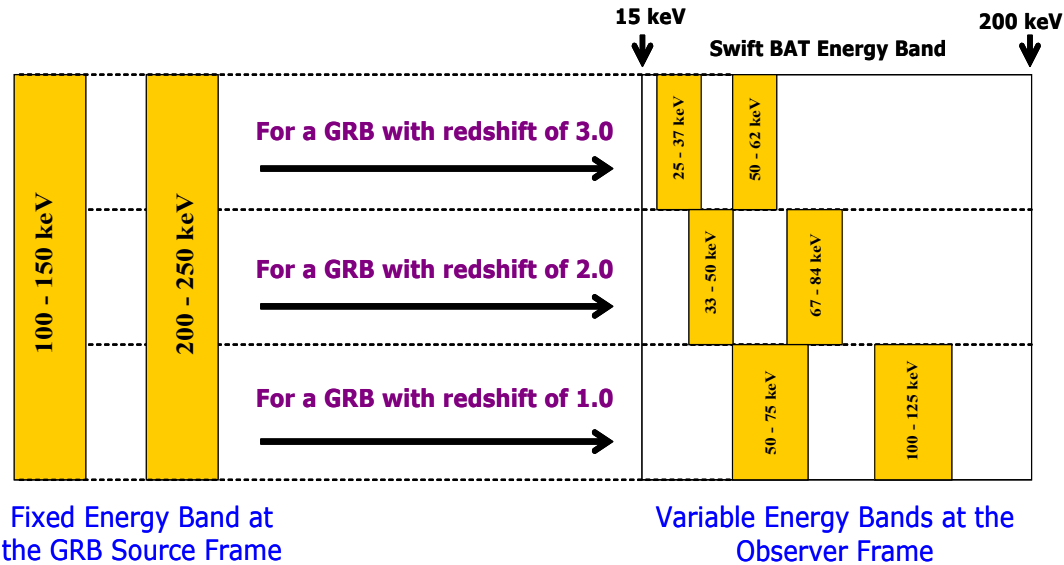}
\caption{Fixed energy bands in the GRB source frame are transformed to various energy bands in the observer frame,
depending on the redshift. Reproduced from Ref. \cite{2012MNRAS.419..614U}.}
\label{fig:LIV2}
\end{figure*}

It should be noted that there are two limitations in the treatment of Ellis et al. \cite{2006APh....25..402E}.
First, they extracted spectral lags in the light curves between two fixed observer-frame energy bands.
However, because different GRBs have different redshift measurements, these two energy bands correspond to
a different pair of energy bands in the source frame \cite{2012MNRAS.419..614U}, thus potentially causing
an artificial energy dependence to the extracted spectral lag and/or a systematic uncertainty to the search for LIV lags.
\footnote{Due to the redshift dependence of cosmological sources, observer frame quantities can be quite different
than source frame ones. In principle, there is a similar problem associated with the energy dependence of the observer-frame
quantities for vacuum birefringence or WEP bounds (more on this below) when analyzing a large sample of cosmological sources
with different redshifts. However, note that if we work on the observer-frame quantities of individual
cosmological sources, there is no such a problem.}
Ukwatta et al. \cite{2012MNRAS.419..614U} found that there is a large scatter in the correlation between observer-frame
lags and source-frame lags for the same GRB sample, indicating that the observer-frame lag does not faithfully
represent the source-frame lag. The first limitation of Ellis et al. treatment can be resolved by selecting
two appropriate energy bands fixed in the rest frame and calculating the time lag for two projected observer-frame
energy bands by the relation $E_{\rm observer}=E_{\rm source}/(1+z)$. Bernardini et al. \cite{2015MNRAS.446.1129B}
studied the source-frame spectral lags of 56 GRBs detected by Swift/Burst Alert Telescope (BAT). For each GRB,
they extracted light curves for two observer-frame energy bands corresponding to the fixed energy bands in
the source frame, i.e., 100--150 and 200-250 keV. These two particular source-frame energy bands were selected
to ensure that the projected observer-frame energy bands (i.e., $[100-150]/(1+z)$ and $[200-250]/(1+z)$ keV) are
within the detectable energy range of the BAT instrument (see Figure~\ref{fig:LIV2}). For each extracted light-curve pairs,
they used the discrete cross-correlation function to calculate the spectral lag. Note that the energy difference
between the median-values of the two source-frame energy bands is fixed at 100 keV, whereas in the observer frame,
the energy difference varies depending on the redshift of each burst. This is in contrast to the spectral lag
extractions achieved in the observer frame, where the energy difference is treated as a constant \cite{2012MNRAS.419..614U}.
Wei \& Wu \cite{2017ApJ...851..127W} first took advantage of the source-frame spectral lags of 56 Swift GRBs
presented in Bernardini et al. \cite{2015MNRAS.446.1129B} to investigate the LIV effects.
For the subluminal case, the arrival-time difference of two photons with observer-frame energy difference $\Delta E$
that induced by the linear LIV reads:
\begin{equation}
\begin{aligned}
\Delta t_{\rm LIV}&=\frac{\Delta E}{H_{0}E_{\rm QG}}\int_{0}^{z}\frac{(1+z'){\rm d}z'}{h(z')}\\
                  &=\frac{\Delta E'/(1+z)}{H_{0}E_{\rm QG}}\int_{0}^{z}\frac{(1+z'){\rm d}z'}{h(z')}\;,
\end{aligned}
\end{equation}
where $\Delta E'=100~\mathrm{keV}$ is the rest-frame energy difference. Similar to Ellis et al. \cite{2006APh....25..402E},
one can formulate the intrinsic time lag problem in terms of linear regression:
\begin{equation}
\frac{\Delta t_{\rm src}}{1+z}=a'_{\rm LV}K'+b_{\rm sf}\;,
\end{equation}
where $\Delta t_{\rm src}$ is the extracted spectral lag for the source-frame energy bands 100--150 and 200-250 keV,
\begin{equation}
K'=\frac{1}{(1+z)^{2}}\int_{0}^{z}\frac{(1+z')dz'}{h(z')}
\end{equation}
is a dimensionless redshift function, and $a'_{\rm LV}=\Delta E'/(H_0 E_{\rm QG})$ is the slope in $K'$.
Using the sample of 56 GRBs with known redshifts, Wei \& Wu \cite{2017ApJ...851..127W} obtained robust
limits on the slope $a'_{\rm LV}$ and the intercept $b_{\rm sf}$ by fitting their source-frame spectral lag data.
The 95\% confidence-level lower limit on the QG energy scale derived from $a'_{\rm LV}$ is
$E_{\rm QG}\geq2.0\times10^{14}$ GeV \cite{2017ApJ...851..127W}. This is a step forward in the study of LIV effects,
since all previous investigations used spectral lags extracted in the observer frame only.

\begin{figure}
\vskip-0.2in
\centerline{\includegraphics[angle=0,width=0.6\textwidth]{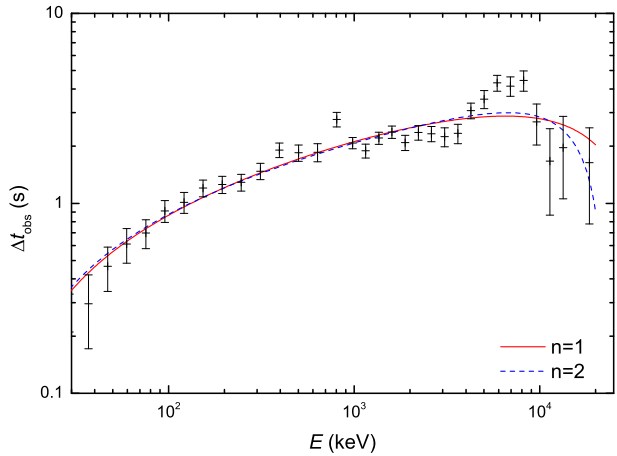}}
\vskip-0.2in
\caption{Energy dependence of the observed time lag $\Delta t_{\rm obs}$ between the lowest-energy band and any other
high-energy bands of GRB 160625B. The solid and dashed curves correspond to the best-fit linear ($n=1$) and quadratic
($n=2$) LIV models, respectively. Reproduced from Ref. \cite{2017ApJ...834L..13W}.}
\label{fig:LIV3}
\end{figure}

The second limitation of Ellis et al. treatment is that an unknown constant was supposed to be the intrinsic time delay
in the linear fitting function (see Eq.~\ref{eq:LIV-linear}), which is tantamount to suggesting that all GRBs have
the same intrinsic time lag. However, since the time durations of GRBs span nearly six orders of magnitude,
it is not possible that high-energy photons radiated from different GRBs (or from the same burst) have the same
intrinsic time lag relative to the radiation time of low-energy photons \cite{2016ChPhC..40d5102C}. As an improvement,
Zhang \& Ma \cite{2015APh....61..108Z} fitted the data of the high-energy photons from GRBs on several straight lines
with the same slope as $1/E_{{\rm QG},n}^{n}$ but with different intercepts (i.e., different intrinsic emission times;
see also Refs.~\cite{2016APh....82...72X,2016PhLB..760..602X,2018JCAP...01..050X}).
However, photons from different GRBs fall on the same line, which still implies that the intrinsic time lags between
the high-energy photons and the low-energy (trigger) photons are much the same for these GRBs.
Chang et al.~\cite{2012APh....36...47C} made use of the magnetic jet model to estimate the intrinsic emission time delay
between high- and low-energy photons from GRBs. However, the magnetic jet model depends on some specific theoretical
parameters, and thus introduces uncertainties on the LIV results. In 2017, Wei et al.~\cite{2017ApJ...834L..13W}
first proposed that GRB 160625B, the burst having a well-defined transition from positive to negative spectral lags
(see Figure~\ref{fig:LIV3}), provides a good opportunity not only to disentangle the intrinsic time
lag problem but also to put new constraints on LIV. The spectral lag is conventionally defined positive
when high-energy photons arrive earlier than low-energy photons, while a negative lag corresponds to a delayed arrival of
high-energy photons. As discussed above, the LIV-induced time delay $\Delta t_{\rm LIV}$ is likely to be accompanied by
a potential intrinsic energy-dependent time lag $\Delta t_{\rm int}$ due to unknown properties of the source.
Therefore, the observed time lag between two different energy bands of a GRB should consist of two parts,
\begin{equation}
\Delta t_{\rm obs}=\Delta t_{\rm int} + \Delta t_{\rm LIV} \;.
\end{equation}
Since the observed time lags of most GRBs have a positive energy dependence (e.g.,
Refs.~\cite{2017ApJ...844..126S,2018ApJ...865..153L}), Wei et al.~\cite{2017ApJ...834L..13W} approximated the observer-frame
relation of the intrinsic time lag and the energy $E$ as a power law with positive dependence,
\begin{equation}
\Delta t_{\rm int}(E)=\tau\left[\left(\frac{E}{\rm keV}\right)^{\alpha}-\left(\frac{E_{0}}{\rm keV}\right)^{\alpha}\right]\;{\rm s} \;,
\end{equation}
where $\tau>0$ and $\alpha>0$, and where $E_{0}=11.34$ keV is the median value of the lowest reference energy band
(10--12 keV). We emphasize that the intrinsic positive time lag implies an earlier arrival time for the higher energy photons.
Also, when the subluminal case ($s_{\pm}=+1$) is considered, high-energy photons would arrive on Earth after low-energy ones,
implying a negative spectral lag due to the LIV effects. As the Lorentz-violating term becomes dominant at higher energy
scales, the positive correlation between the time lag and the energy would gradually trend in an opposite way.
The combined contributions from the intrinsic time lag and the LIV-induced time lag can therefore produce the observed lag
behavior with a turnover from positive to negative lags \cite{2017ApJ...834L..13W}. By fitting the spectral lag behavior
of GRB 160625B, Wei et al.~\cite{2017ApJ...834L..13W,2017ApJ...842..115W} obtained both a reasonable formulation of
the intrinsic energy-dependent time lag and comparatively robust limits on the QG energy scale and the coefficients
of the Standard Model Extension. The $1\sigma$ confidence-level lower limits are
$E_{{\rm QG},1}>0.5\times10^{16}~\mathrm{GeV}$ and $E_{{\rm QG}, 2}>1.4\times10^{7}~\mathrm{GeV}$
for linear and quadratic LIV, respectively. The spectral lag data of GRB 160625B does not have the current best
sensitivity to LIV constraints, but the analysis method, when applied to future bright short GRBs with similar
spectral-lag transitions, may result in more stringent constraints on LIV.

\begin{itemize}
  \item Active galactic nuclei
\end{itemize}

Thanks to their fast flux variations (hours to years), cosmological distances, and very high energy
(VHE, $E\geq100~\mathrm{GeV}$)
gamma-rays, TeV flares of AGNs have also been viewed as very effective probes for searching for the LIV-induced
vacuum dispersions. It is worth pointing out that testing LIV with both GRBs and flaring AGNs is of great fundamental
interest. GRBs can be detected at very large distances (up to $z\sim8$), but with very limited high-energy
($E>$ tens of GeV) photons. On the contrary, AGN flares can be well observed with large statistics of photons
up to a few tens of TeV. But due to extinction of high-energy photons by extragalactic background light,
TeV detections are limited to those sources with relatively low redshifts $z\leq0.5$. Hence, GRBs and flaring AGNs are
mutually complementary in testing LIV, and they allow to test different redshift and energy ranges. There have
been some resulting constraints on LIV using TeV observations of bright AGN flares, including the Whipple analysis
of the flare of Mrk 421 \cite{1999PhRvL..83.2108B}, the MAGIC and H.E.S.S. analyses of the flares of Mrk 501
\cite{2008PhLB..668..253M,2009APh....31..226M,2019ApJ...870...93A}, and the H.E.S.S. analysis of the flare of
PKS 2155-304 \cite{2008PhRvL.101q0402A,2011APh....34..738H}. The current best limit for the linear case considering
a subluminal LIV effect obtained with AGNs is from the H.E.S.S. analysis of the PKS 2155-304 flare data, namely
$E_{{\rm QG}, 1}>2.1\times10^{18}~\mathrm{GeV}$ \cite{2011APh....34..738H}. For the quadratic LIV, the best
limits derived from AGNs have been set by H.E.S.S.'s observation of the TeV flare of Mrk 501. The reported limits
are $E_{{\rm QG}, 2}>8.5\times10^{10}~\mathrm{GeV}$ ($E_{{\rm QG}, 2}>7.3\times10^{10}~\mathrm{GeV}$) for the subluminal (superluminal) case \cite{2019ApJ...870...93A}.

\begin{itemize}
  \item Pulsars
\end{itemize}

A third class of astrophysical sources used for the time-of-flight tests on gamma-rays are pulsars.
When it comes to testing the quadratic LIV term, having a VHE emission will compensate for a short of distance.
Gamma-ray pulsars, albeit being detected many orders of magnitude closer than GRBs or AGNs, have the advantage of
precisely periodic flux variation, as well as the fact that they are the only stable candidate astrophysical sources
for such time-of-flight studies. Sensitivity to LIV can therefore be improved by simply observing longer.
Additionally, since the timing of the pulsar is carefully studied throughout the electromagnetic spectrum,
energy-dependent time delays induced by propagation effects can be more easily distinguished from intrinsic delays.
First limits on LIV using gamma-ray radiation from the galactic Crab pulsar were obtained from the observation of
the Energetic Gamma-Ray Experiment Telescope (EGRET) onboard the Compton Gamma-Ray Observatory (CGRO) at energies
above 2 GeV \cite{Kaaret1999}, and improved by the VERITAS data above 120 GeV \cite{2011ICRC....7..256O,2013ICRC...33.2768Z}.
Recently, the MAGIC collaboration presented the best limits derived from pulsars by studying the Crab pulsar emission
observed up to TeV energies, yielding $E_{{\rm QG}, 1}>5.5\times10^{17}~\mathrm{GeV}$
($E_{{\rm QG}, 1}>4.5\times10^{17}~\mathrm{GeV}$) for a linear, and $E_{{\rm QG}, 2}>5.9\times10^{10}~\mathrm{GeV}$
($E_{{\rm QG}, 2}>5.3\times10^{10}~\mathrm{GeV}$) for a quadratic LIV, for the subluminal (superluminal) case, respectively
\cite{2017ApJS..232....9M}.

The most important constraints obtained so far with vacuum dispersion time-of-flight measurements of various astrophysical
sources are summarized in Table~\ref{table1}. The most stringent lower limits to date on the linear and quadratic LIV
energy scales were set by the observation of GRB 090510 with Fermi/LAT. The values for the subluminal (superluminal) case
are $E_{{\rm QG}, 1}>9.3\times10^{19}~\mathrm{GeV}$ ($E_{{\rm QG}, 1}>1.3\times10^{20}~\mathrm{GeV}$) and
$E_{{\rm QG}, 2}>1.3\times10^{11}~\mathrm{GeV}$ ($E_{{\rm QG}, 2}>9.4\times10^{10}~\mathrm{GeV}$) \cite{2013PhRvD..87l2001V}.
Clearly, these vacuum dispersion studies using gamma rays in the GeV--TeV range offer us
at present with the best opportunity to search for Planck-scale modifications of the dispersion
relation. Unfortunately, while they provide meaningful bounds for the linear ($n=1$) modification,
they are much weaker for deviations that arise at the quadratic ($n=2$) order.

\begin{table*}
\caption{A selection of lower limits on $E_{{\rm QG}, n}$ for linear ($n=1$) and quadratic ($n=2$) LIV for the subluminal ($s_{\pm}=+1$) and superluminal ($s_{\pm}=-1$) cases. These limits were obtained from vacuum dispersion time-of-flight measurements of various astrophysical sources.}
\scriptsize
\begin{tabular}{lllccccl}
\hline
\hline
Source(s) & Instrument & Technique &   \multicolumn{2}{c}{$E_{{\rm QG}, 1}$ (GeV)}    & \multicolumn{2}{c}{$E_{{\rm QG}, 2}$ (GeV)}    & Refs. \\
\cmidrule(lr){4-5}\cmidrule(lr){6-7}
          &            &           & $s_{\pm}=+1$  &    $s_{\pm}=-1$  & $s_{\pm}=+1$   &    $s_{\pm}=-1$  &              \\
\hline
9 GRBs$^{a}$  &  BATSE+OSSE  &  Wavelets  & $6.9\times10^{15}$ & --- & $2.9\times10^{6}$ & --- &\cite{Ellis2003}  \\
GRB 021206$^{b}$  &  RHESSI  &  Peak times at different energies & $1.8\times10^{17}$ & --- & --- & --- & \cite{2004ApJ...611L..77B}  \\
35 GRBs$^{c}$  &  BATSE+HETE-2  &  Wavelets  & $1.4\times10^{16}$ & --- & --- & --- & \cite{2006APh....25..402E,2008APh....29..158E}  \\
      & +Swift  &   &   &  &  &  &   \\
GRB 051221A  &  Konus-Wind+Swift  &  Peak times of the light curves  & $6.6\times10^{16}$ & --- & $6.2\times10^{6}$ & --- & \cite{2006JCAP...05..017R}   \\
      &   & in different energy bands  &   &  &  &  &  \\
15 GRBs  &  HETE-2  &  Wavelets  & $2.0\times10^{15}$ & --- & --- & --- & \cite{2008ApJ...676..532B}  \\
11 GRBs  &  INTEGRAL  &  Likelihood  & $3.2\times10^{11}$ & --- & --- & --- & \cite{2008GReGr..40.1731L}   \\
GRB 080916C  &  Fermi GBM+LAT  &  Associating a 13.2 GeV photon with   & $1.3\times10^{18}$ & --- & --- & --- & \cite{2009Sci...323.1688A}   \\
      &   &  the trigger time &   &  &  & &  \\
GRB 090510  &  Fermi GBM+LAT  &  Associating a 31 GeV photon with   & $1.5\times10^{19}$ & --- & --- & --- & \cite{2009Natur.462..331A}   \\
      &   &  the start of the first GBM pulse &   &  &  & &  \\
   &  Fermi/LAT  &  PairView+Likelihood  & $9.3\times10^{19}$ & $1.3\times10^{20}$ & $1.3\times10^{11}$ & $9.4\times10^{10}$ & \cite{2013PhRvD..87l2001V}   \\
      &   &  +Sharpness-Maximization Method &   &  &  & & \\
GRB 160625B  &  Fermi/GBM  &  Spectral lag transition & $0.5\times10^{16}$ & --- & $1.4\times10^{7}$ & --- & \cite{2017ApJ...834L..13W}   \\
56 GRBs  &  Swift &  Rest-frame spectral lags & $2.0\times10^{14}$ & --- & --- & --- & \cite{2017ApJ...851..127W}  \\
GRB 190114C  &  MAGIC  &  Likelihood  & $5.8\times10^{18}$ & $5.5\times10^{18}$ & $6.3\times10^{10}$ & $5.6\times10^{10}$ & \cite{PhysRevLett.125.021301}   \\
\hline
Mrk 421  &  Whipple  &  Binning  & $4.0\times10^{16}$ & --- & --- & --- & \cite{1999PhRvL..83.2108B}   \\
Mrk 501  &  MAGIC &  Energy cost function  & $2.1\times10^{17}$ & --- & $2.6\times10^{10}$ & --- & \cite{2008PhLB..668..253M}   \\
         &        &  Likelihood  & $3.0\times10^{17}$ & --- & $5.7\times10^{10}$ & --- & \cite{2009APh....31..226M}  \\
         &  H.E.S.S. &  Likelihood  & $3.6\times10^{17}$ & $2.6\times10^{17}$ & $8.5\times10^{10}$ & $7.3\times10^{10}$ & \cite{2019ApJ...870...93A}   \\
PKS 2155-304 &  H.E.S.S. &  Modified cross correlation function  & $7.2\times10^{17}$ & --- & $1.4\times10^{9}$ & --- & \cite{2008PhRvL.101q0402A}   \\
             &           &  Likelihood  & $2.1\times10^{18}$ & --- & $6.4\times10^{10}$ & --- & \cite{2011APh....34..738H}  \\
\hline
Crab pulsar  &  CGRO/EGRET  &  Pulse arrival times   & $1.8\times10^{15}$ & --- & --- & --- & \cite{Kaaret1999}   \\
      &   & in different energy bands  &   &  &  &  &  \\
             &  VERITAS  &  Likelihood    & $3.0\times10^{17}$ & --- & $7.0\times10^{9}$  & --- & \cite{2011ICRC....7..256O}   \\
             &           &  Dispersion Cancellation    & $1.9\times10^{17}$ & $1.7\times10^{17}$ & --- & --- & \cite{2013ICRC...33.2768Z}   \\
             &  MAGIC  &  Likelihood    & $5.5\times10^{17}$ & $4.5\times10^{17}$ & $5.9\times10^{10}$  & $5.3\times10^{10}$ & \cite{2017ApJS..232....9M}   \\
\hline
\end{tabular}
\label{table1}
\\
$^{a}$Limits obtained not taking into account the factor $(1+z)$ in the intergrand of Eq.~(\ref{eq:tLIV}).
\\
$^{b}$The pseudo redshift was estimated from the spectral and temporal properties of GRB 021206.
\\
$^{c}$The Limits of Ellis et al. \cite{2006APh....25..402E} were corrected in Ellis et al. \cite{2008APh....29..158E} taking into account the factor $(1+z)$ in the intergrand of Eq.~(\ref{eq:tLIV}).
\end{table*}

\subsection{Vacuum birefringence from LIV}\label{subsec:LIV2}
\subsubsection{General formulae}
In QG theories that invoke LIV, the Charge-Parity-Time (CPT) theorem, i.e., the invariance of the laws of physics under
charge conjugation, parity transformation, and time reversal, no longer holds. Note that the fact CPT does not
hold does not imply that it should be violated. In the absence of Lorentz invariance, the CPT invariance, if needed, should be
imposed as an additional assumption. In the effective field theory approach \cite{2003PhRvL..90u1601M}, the Lorentz- and
CPT-violating dispersion relation for photon propagation can be parameterized as
\begin{equation}\label{eq:dispersion}
  E_{\pm}^2=p^2c^2\pm \frac{2\eta}{E_{\rm pl}} p^3c^3\;,
\end{equation}
where $\pm$ represents the left- or right-handed circular polarization states of the photon, and $\eta$ is a dimensionless
parameter that needs to be constrained. In LIV but CPT invariant theories, the parameter $\eta$ exactly vanishes. In this sense,
such tests might be less general than the ones based on vacuum dispersion.
The linear polarization can be decomposed into left- and right-handed
circular polarization states. For $\eta\neq0$, photons with opposite circular polarizations have
slightly different group velocities, which leads to a rotation of the polarization vector of a linearly polarized wave.
This effect is known as vacuum birefringence. The rotation angle propagating from the source at redshift $z$ to the observer
can be derived as \cite{2011PhRvD..83l1301L,2012PhRvL.109x1104T}
\begin{equation}\label{eq:theta-LIV}
  \Delta\phi_{\rm LIV}(E)\simeq\eta\frac{E^2 F(z)}{\hbar E_{\rm pl}H_{0}}\;,
\end{equation}
where $E$ is the observed photon energy, and
\begin{equation}
F(z)=\int_0^z\frac{\left(1+z'\right)dz'}{\sqrt{\Omega_{\rm m}\left(1+z'\right)^3+\Omega_{\Lambda}}}\;.
\end{equation}

Generally speaking, it is impossible to know the intrinsic polarization angles in the emission of photons of different
energies from a given source. If one had this information, evidence for vacuum birefringence (i.e., an energy-dependent
rotation of the polarization plane) can be examined by measuring differences between the known intrinsic polarization angle
and the observed polarization angles at different energies. However, even without such knowledge, the birefringent effect
can still be constrained for polarized sources at arbitrary cosmological distances, because the differential rotation acting
on the polarization angle as a function of energy would add opposite oriented polarization vectors, effectively erasing
most, if not all, of the observed polarization signal. Therefore, the detection of the polarization signal can put an
upper bound on such a possible violation.

\subsubsection{Present constraints}
Observations of linear polarization from distant sources have been widely used to place upper limits on the
birefringent parameter $\eta$. The vacuum birefringence constraints arise from the fact if the rotation angle
(Eq.~\ref{eq:theta-LIV}) differs by more than $\pi/2$ over an energy range ($E_1<E<E_2$), then the net polarization
of the signal would be severely suppressed and well below any observed value. Hence, the measurement of polarization
in a given energy band implies that the differential rotation angle $|\Delta\phi(E_{2})-\Delta\phi(E_{1})|$
should not be larger than $\pi/2$.

Previously, Gleiser \& Kozameh \cite{2001PhRvD..64h3007G} set an upper bound of $\eta<10^{-4}$ by analyzing
the linearly polarized ultraviolet light from the distant radio galaxy 3C 256. Much more stringent limits, $\eta<10^{-14}$,
have been
obtained by using the linear polarization detection in the gamma-ray emission of GRB 021206
\cite{2003Natur.426Q.139M,2004PhRvL..93b1101J}. However, the originally reported detection of high polarization from GRB 021206
\cite{2003Natur.423..415C} has been refuted by the re-analyses of the same data \cite{2004MNRAS.350.1288R,2004ApJ...613.1088W}.
Maccione et al. \cite{2008PhRvD..78j3003M} used the hard X-ray polarization measurement of the Crab Nebula
to get a constraint of $\eta<9\times10^{-10}$.
Laurent et al.~\cite{2011PhRvD..83l1301L} used a report of polarized soft gamma-ray emission from GRB 041219A to
derive a stronger limit of $\eta<1.1\times10^{-14}$ (see also Ref.~\cite{2011APh....35...95S}). But again,
this claimed polarization detection \cite{2007ApJS..169...75K,2007A&A...466..895M,2009ApJ...695L.208G} has been disputed
(see the explanations in Ref.~\cite{2012PhRvL.109x1104T}). That is, the previous reports of the gamma-ray polarimetry
for GRB 041219A are controversial and, thus, the arguments for the limits on $\eta$ given by
Refs.~\cite{2011PhRvD..83l1301L,2011APh....35...95S} are still open to questions.

Contrary to those disputed reports, the evidences of linearly polarized gamma-ray emission detected by the gamma-ray burst
polarimeter (GAP) onboard the Interplanetary Kite-craft Accelerated by Radiation Of the Sun (IKAROS) are convincing,
and thus these detections can be used to set more reliable limits on the birefringent parameter \cite{2012PhRvL.109x1104T}.
IKAROS/GAP detected gamma-ray polarizations of three GRBs with high significance levels, with a linear polarization degree
of $\Pi=27\pm11\%$ for GRB 100826A \cite{2011ApJ...743L..30Y}, $\Pi=70\pm22\%$ for GRB 110301A, and $\Pi=84^{+16}_{-28}\%$
for GRB 110721A \cite{2012ApJ...758L...1Y}. The detection significance are $2.9\sigma$, $3.7\sigma$, and $3.3\sigma$,
respectively. Toma et al. \cite{2012PhRvL.109x1104T} set the upper limit of the differential rotation angle
$|\Delta\phi(E_{2})-\Delta\phi(E_{1})|$ to be $\pi/2$, and obtained a severe upper limit on $\eta$ in the order of
$\mathcal{O}(10^{-15})$ from the reliable polarimetric data of these three GRBs. However, the GRBs had no direct redshift
measurement, they used a redshift estimate based on an empirical luminosity relation.
Utilizing the real redshift measurement ($z=1.33$) together with the polarization data of GRB 061122,
G{\"o}tz et al. \cite{2013MNRAS.431.3550G} obtained a stricter limit ($\eta<3.4\times10^{-16}$) on a possible LIV.
G{\"o}tz et al. \cite{2014MNRAS.444.2776G} used the most distant polarized burst (up to $z=2.739$), GRB 140206A, to
obtain the deepest limit to date ($\eta<1.0\times10^{-16}$) on the possibility of LIV.

It is worth noting that most of previous polarization constraints were derived under the assumption that
the differential rotation angle $|\Delta\phi(E_{2})-\Delta\phi(E_{1})|$ is smaller than $\pi/2$.
However, Lin et al. \cite{2016MNRAS.463..375L} gave a detailed analysis on the evolution of GRB
polarization arising from the vacuum birefringent effect, and showed that a considerable amount of
the initial polarization degree (depending both on the photon energy band and the photon spectrum)
can be conserved even if $|\Delta\phi(E_{2})-\Delta\phi(E_{1})|$ is approaching to $\pi/2$.
This is incompatible with the common belief that $|\Delta\phi(E_{2})-\Delta\phi(E_{1})|$ should not be
too large when high polarization is detected. Therefore, Lin et al. \cite{2016MNRAS.463..375L} suggested that
it is unsuitable to constrain the birefringent effect by simply setting $\pi/2$ as the upper limit of
$|\Delta\phi(E_{2})-\Delta\phi(E_{1})|$. Lin et al. \cite{2016MNRAS.463..375L} applied their formulae for
the polarization evolution to some true GRB events, and obtained the most stringent upper limit to date
on the birefringent parameter from the polarimetric data of GRB 061122, i.e., $\eta<0.5\times10^{-16}$.
Following the analysis method presented in Ref.~\cite{2016MNRAS.463..375L} and utilizing the recent
measurements of gamma-ray linear polarization of GRBs, Wei \cite{2019MNRAS.485.2401W} updated constraints
on a possible LIV through the vacuum birefringent effect, and thereby improving previous results by
factors ranging from two to ten.

\begin{figure*}
\centering
\vskip-0.2in
\includegraphics[width=0.7\textwidth]{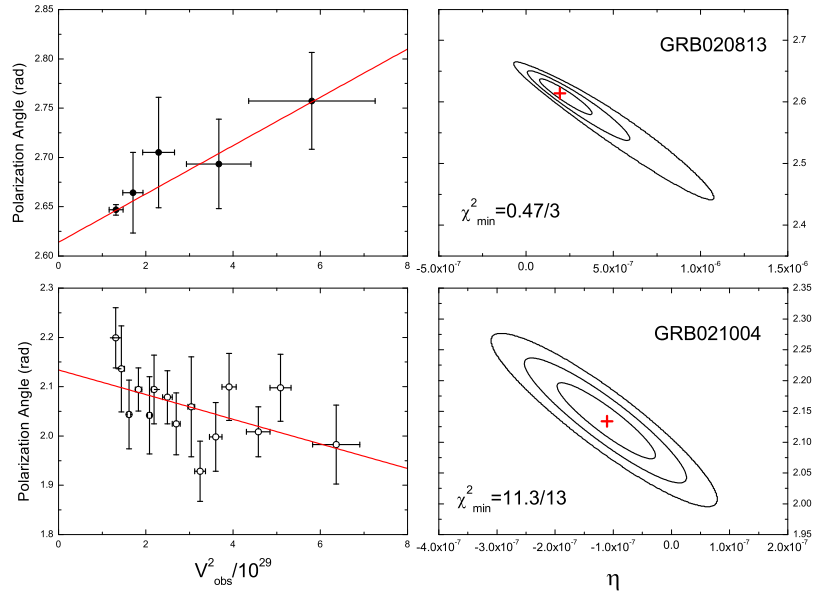}
\vskip-0.2in
\caption{Fit to the multiwavelength polarimetric data of the optical afterglows of GRB 020813 and GRB 021004.
Left panels: linear fits of the observed linear polarization angles versus quantities of frequency squared.
Right panels: 1-3$\sigma$ confidence levels in the $\phi_{0}$-$\eta$ plane. Reproduced from Ref. \cite{2007MNRAS.376.1857F}.}
\label{fig:LIV4}
\end{figure*}

\begin{table*}
\caption{A selection of limits on the vacuum birefringent parameter $\eta$ from the linear polarization measurements
of astrophysical sources.}
\scriptsize
\begin{tabular}{lcccccc}
\hline
\hline
Author (year) & Source & Polarimeter & Energy band$^{a}$ &  $\Pi$ & $\eta$    & Refs. \\
\hline
Gleiser and Kozameh (2001) & 3C 256 & Spectropolarimeter & Ultraviolet & $16.4\pm2.2\%$ & $<10^{-4}$ & \cite{2001PhRvD..64h3007G} \\
Mitrofanov (2003) & GRB 021206$^{c}$ & RHESSI & 150--2000 keV & $80\pm20\%$$^{b}$ & $<10^{-14}$ & \cite{2003Natur.426Q.139M} \\
Jacobson et al. (2004) & GRB 021206$^{d}$ & RHESSI & 150--2000 keV & $80\pm20\%$$^{b}$ & $<0.5\times10^{-14}$ & \cite{2004PhRvL..93b1101J} \\
Fan et al. (2007) & GRB 020813 & LRISp & 3500--8800 {\AA} & 1.8\%--2.4\% & ($-2.0$--1.4)$\times10^{-7}$ & \cite{2007MNRAS.376.1857F} \\
                  & GRB 021004 & VLT & 3500--8600 {\AA} & $1.88\pm0.05\%$ & ($-2.0$--1.4)$\times10^{-7}$ & \\
Maccione et al. (2008) & Crab Nebula & INTEGRAL/SPI & 100--1000 keV & $46\pm10\%$ & $<9\times10^{-10}$ & \cite{2008PhRvD..78j3003M} \\
Laurent et al. (2011) & GRB 041219A$^{e}$ & INTEGRAL/IBIS & 200--800 keV & $43\pm25\%$$^{b}$ & $<1.1\times10^{-14}$ & \cite{2011PhRvD..83l1301L} \\
Stecker (2011) & GRB 041219A$^{f}$ & INTEGRAL/SPI & 100--350 keV & $96\pm40\%$$^{b}$ & $<2.4\times10^{-15}$ & \cite{2011APh....35...95S}\\
Toma et al. (2012) & GRB 100826A$^{f}$ & IKAROS/GAP & 70--300 keV & $27\pm11\%$ & $<2.0\times10^{-14}$ & \cite{2012PhRvL.109x1104T}\\
                   & GRB 110301A$^{f}$ & IKAROS/GAP & 70--300 keV & $70\pm22\%$ & $<1.0\times10^{-14}$ & \\
                   & GRB 110721A$^{f}$ & IKAROS/GAP & 70--300 keV & $84^{+16}_{-28}\%$ & $<2.0\times10^{-15}$ & \\
G{\"o}tz et al. (2013) & GRB 061122 & INTEGRAL/IBIS & 250--800 keV & $>60\%$ & $<3.4\times10^{-16}$ & \cite{2013MNRAS.431.3550G}\\
G{\"o}tz et al. (2014) & GRB 140206A & INTEGRAL/IBIS & 200--400 keV & $>48\%$ & $<1.0\times10^{-16}$ & \cite{2014MNRAS.444.2776G}\\
Lin et al. (2016) & GRB 061122 & INTEGRAL/IBIS & 250--800 keV & $>60\%$ & $<0.5\times10^{-16}$ & \cite{2016MNRAS.463..375L}\\
                  & GRB 110721A$^{f}$ & IKAROS/GAP & 70--300 keV & $84^{+16}_{-28}\%$ & $<4.0\times10^{-16}$ & \\
Wei (2019)   &     GRB 061122	& INTEGRAL/IBIS	&	250--800 keV	&$	>60\%	$	&	$<0.5\times10^{-16}$ & \cite{2019MNRAS.485.2401W}\\
             &     GRB 100826A$^{f}$	& IKAROS/GAP	&	70--300 keV	&$	27\pm11\%	$ &	 $1.2^{+1.4}_{-0.7}\times10^{-14}$ & \\
             &     GRB 110301A$^{f}$	& IKAROS/GAP	&	70--300	keV &$	70\pm22\%	$&	 $4.3^{+5.4}_{-2.3}\times10^{-15}$ & \\
             &     GRB 110721A	& IKAROS/GAP	&	70--300 keV &$	84^{+16}_{-28}\%$&	 $5.1^{+4.0}_{-5.1}\times10^{-16}$ & \\
             &     GRB 140206A	& INTEGRAL/IBIS	&	200--400 keV &$	>48\%	$&	$<1.0\times10^{-16}$ & \\
             &     GRB 160106A$^{f}$	& AstroSat/CZTI	&	100--300 keV	&$	68.5	\pm	24\%$&	 $3.4^{+1.4}_{-1.8}\times10^{-15}$ & \\
             &     GRB 160131A	& AstroSat/CZTI	&	100--300 keV	&$	94	\pm	31\%	$&	 $1.2^{+2.0}_{-1.2}\times10^{-16}$ & \\
             &     GRB 160325A$^{f}$	& AstroSat/CZTI	&	100--300 keV	&$	58.75\pm23.5\%	$&	 $2.3^{+1.0}_{-0.9}\times10^{-15}$ & \\
             &     GRB 160509A	& AstroSat/CZTI	&	100--300 keV	&$	96	\pm	40\%	$&	 $0.8^{+2.2}_{-0.8}\times10^{-16}$ & \\
             &     GRB 160802A$^{f}$	& AstroSat/CZTI	&	100--300 keV	&$	85	\pm	29\%	$&	 $2.0^{+1.7}_{-2.0}\times10^{-15}$ & \\
             &     GRB 160821A$^{f}$	& AstroSat/CZTI	&	100--300 keV	&$	48.7\pm	14.6\%	$&	 $8.9^{+1.7}_{-1.7}\times10^{-15}$ & \\
             &     GRB 160910A$^{f}$	& AstroSat/CZTI	&	100--300 keV	&$	93.7\pm	30.92\%	$&	 $4.7^{+7.6}_{-4.7}\times10^{-16}$ & \\
\hline
\end{tabular}
\label{table2}
\\
$^{a}$The energy band in which polarization is observed.
\\
$^{b}$The claimed polarization detections have been refuted.
\\
$^{c}$The distance of GRB 021206 was taken to be $10^{10}$ light years.
\\
$^{d}$The distance of GRB 021206 was taken to be 0.5 Gpc.
\\
$^{e}$The lower limit to the photometric redshift of GRB 041219A ($z=0.02$) was adopted.
\\
$^{f}$The redshifts of these GRBs were estimated by the empirical luminosity relation.
\end{table*}

If both the intrinsic polarization angle $\phi_{0}$ and the rotation angle induced by the birefringent effect
$\Delta\phi_{\rm LIV}(E)$ are considered here, the observed linear polarization angle at a certain $E$
from a source should be
\begin{equation}
\phi_{\rm obs}=\phi_{0}+\Delta\phi_{\rm LIV}\left(E\right)\;.
\end{equation}
Assuming that all photons in the observed energy bandpass are emitted with the same (unknown) intrinsic polarization
angle, we then expect to observe the birefringent effect as an energy-dependent linear polarization vector.
To constrain the birefringent parameter $\eta$, Fan et al. \cite{2007MNRAS.376.1857F} looked for a similar
energy-dependent trend in the multiwavelength polarization observations of the optical afterglows of GRB 020813
and GRB 021004. By fitting the spectropolarimetric data of these two GRBs, Fan et al. \cite{2007MNRAS.376.1857F}
obtained constraints on both $\phi_{0}$ and $\eta$ (see Figure~\ref{fig:LIV4}). At the $3\sigma$ confidence level,
the combined limit on $\eta$ from two GRBs is $-2\times10^{-7}<\eta<1.4\times10^{-7}$.
It is clear from Eq.~(\ref{eq:theta-LIV}) that the higher energy band of the polarization observation
and the larger distance of the polarized source, the greater sensitivity to small values of $\eta$.
As expected, the optical polarization data of GRB afterglow obtained a less stringent constraint on $\eta$
\cite{2007MNRAS.376.1857F}.

\subsubsection{Comparison with time-of-flight limits}

Table~\ref{table2} presents a summary of the corresponding limits on LIV from the polarization measurements of
astrophysical sources. The hitherto most stringent constraints on the birefringent parameter,
$\eta<\mathcal{O}(10^{-16})$, have been obtained by the detections of linear polarization in the prompt gamma-ray
emission of GRBs \cite{2013MNRAS.431.3550G,2014MNRAS.444.2776G,2016MNRAS.463..375L,2019MNRAS.485.2401W}.

By comparing Eqs.~(\ref{eq:LIVdispersion}) and (\ref{eq:dispersion}), we can derive the conversion from $\eta$
to the Limit on the linear LIV energy scale, i.e., $E_{{\rm QG}, 1}=\frac{E_{\rm pl}}{2\eta}$.
With the data presented in Tables~\ref{table1} and \ref{table2}, it is easy to compare the recent achievements
in sensitivity of time-of-flight measurements versus polarization measurements. For the subluminal case,
the time-of-flight analysis of multi-GeV photons from GRB 090510A detected by Fermi/LAT yielded the strictest limit
on the linear LIV energy scale, $E_{{\rm QG}, 1}>9.3\times10^{19}~\mathrm{GeV}$, which corresponds to $\eta<0.07$
\cite{2013PhRvD..87l2001V}. Obviously, this time-of-flight constraint is many orders of magnitude weaker than
the best polarization constraint. However, time-of-flight constraints are essential in a broad-based search for
nonbirefringent Lorentz-violating effects.

\section{
Astrophysical bounds on the photon mass}\label{sec:photonmass}

\subsection{Dispersion from a nonzero photon mass}\label{subsec:photonmass1}
For a nonzero photon rest mass ($m_\gamma\neq0$), the energy of the photon can be written as
\begin{equation}
E=\sqrt{p^2c^2+m_\gamma^2c^4}\;.
\end{equation}
Then the massive photon group velocity $\upsilon$ in vacuum is no longer a constant $c$,
but depends on the photon frequency $\nu$. That is,
\begin{equation}\label{eq:mr_v}
\upsilon=\frac{\partial{E}}{\partial{p}}=c\sqrt{1-\frac{m_\gamma^2c^4}{E^2}}\approx c\left(1-\frac{1}{2}\frac{m_\gamma^2c^4}{h^2\nu^2}\right)\;,
\end{equation}
where the last approximation is valid when
$m_\gamma\ll h\nu/c^{2}\simeq7\times10^{-42}\left(\frac{\nu}{\rm GHz}\right)\;{\rm kg}$.
Eq.~(\ref{eq:mr_v}) implies that the lower frequency, the slower the photon travels in vacuum.
Two massive photons with different frequencies ($\nu_{l}<\nu_{h}$), if emitted simultaneously
from a same source, would be received at different times by the observer. For a cosmic source,
the arrival time difference due to a nonzero photon mass is given by
\begin{equation}\label{eq:tmr}
  \Delta{t_{m_{\gamma}}}=\frac{m_\gamma^{2} c^4}{2h^{2}H_0}\left(\nu_l^{-2}-\nu_h^{-2}\right)H_{\gamma}(z)\;,
\end{equation}
where $H_{\gamma}(z)$ is a dimensionless function of the source redshift $z$,
\begin{equation}\label{eq:Hr}
  H_{\gamma}(z)=\int_{0}^{z}\frac{(1+z')^{-2}dz'}{\sqrt{\Omega_{\rm m}(1+z')^{3}+\Omega_{\Lambda}}}\;.
\end{equation}
It is obvious from Eq.~(\ref{eq:tmr}) that observations of shorter time structures at lower frequencies
from sources at cosmological distances are particularly powerful for constraining the photon mass.
In contrast, observations of high-energy photons from cosmological sources are better suited for
probing LIV \cite{1998Natur.393..763A}.

\begin{figure}[h]
\vskip-0.2in
\includegraphics[angle=0,scale=0.5]{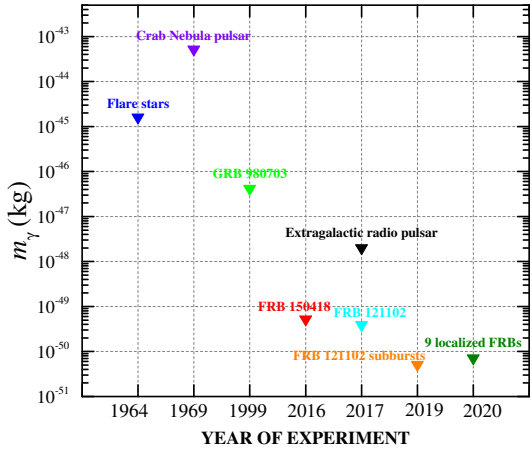}
\vskip-2.4in
\caption{Astrophysical limits on the photon rest mass through the dispersion method, including
the strict upper bounds from flare stars \cite{1964Natur.202..377L}, Crab nebula pulsar \cite{1969Natur.222..157W},
GRB 980703 \cite{1999PhRvL..82.4964S}, extragalactic radio pulsar \cite{2017RAA....17...13W}, FRB 150418
\cite{2016ApJ...822L..15W,2016PhLB..757..548B}, FRB 121102 \cite{2017PhLB..768..326B}, FRB 121102 subbursts
\cite{2019ApJ...882L..13X}, and nine localized FRBs \cite{2020arXiv200609680W}.}
\label{fig:mass1}
\end{figure}

The observed time delays between different energy bands from astrophysical sources have been used to
constrain the photon mass. For instance, Lovell et al.~\cite{1964Natur.202..377L} analysed the delay
between the optical and radio emission from flare stars and concluded that the relative velocity of
light and radio waves was constrained to be $4\times10^{-7}$ over a wavelength range from 0.54 $\mu$m
to 1.2 m, which implied an upper limit on the photon mass of $m_\gamma\leq1.6\times10^{-45}~\mathrm{kg}$.
With the arrival time delay of optical pulses from the Crab Nebula pulsar over a wavelength range of
0.35--0.55 $\mu$m, Warner \& Nather \cite{1969Natur.222..157W} set a stringent limit on the possible
frequency-dependence of the speed of light, which led to an upper limit of $m_\gamma\leq5.2\times10^{-44}~\mathrm{kg}$.
By analyzing the arrival time delay between the radio afterglow and the prompt gamma-ray emission
from GRB 980703, Schaefer \cite{1999PhRvL..82.4964S} obtained a stricter upper limit of
$m_\gamma\leq4.2\times10^{-47}~\mathrm{kg}$. Using GRB early-time radio detections as well as multi-band
radio afterglow peaks, Zhang et al. \cite{2016JHEAp..11...20Z} improved the results of Schaefer \cite{1999PhRvL..82.4964S}
by nearly half an order of magnitude. Although the optical emissions of the Crab Nebula pulsar have been
used to constrain the photon mass, Wei et al. \cite{2017RAA....17...13W} showed that much more severe
limits on the photon mass can be obtained with radio observations of pulsars in the Large and Small
Magellanic Clouds (LMC and SMC). The photon mass limits can be as low as $m_\gamma\leq2.0\times10^{-48}~\mathrm{kg}$
for the radio pulsar PSR J0451-67 in the LMC and $m_\gamma\leq2.3\times10^{-48}~\mathrm{kg}$ for PSR J0045-7042
in the SMC \cite{2017RAA....17...13W}. Owing to their fine time structures, low frequency emissions,
and large cosmological distances, extragalactic fast radio bursts (FRBs) have been viewed as the most
promising celestial laboratory so far for testing the photon mass
\cite{2016ApJ...822L..15W,2016PhLB..757..548B,2017PhLB..768..326B,2017PhRvD..95l3010S,2019ApJ...882L..13X,2020arXiv200609680W}.
The first attempts to constrain the photon mass using FRBs were presented in Wu et al. \cite{2016ApJ...822L..15W}
and Bonetti et al. \cite{2016PhLB..757..548B}. Adopting the possible redshift $z=0.492$ for FRB 150418,
and assuming the dispersive delay was caused by the nonzero photon mass effect,
Wu et al. \cite{2016ApJ...822L..15W} improved the limit of the photon mass to be
$m_{\gamma}\leq5.2\times10^{-50}~\mathrm{kg}$ (see also \cite{2016PhLB..757..548B}).
However, the identification of a radio transient, which provided the redshift measurement to FRB 150418,
was challenged with a common AGN variability \cite{2016ApJ...824L...9V,2016ApJ...821L..22W}. Now
this redshift measurement is generally thought to be unreliable \cite{2017Natur.541...58C}.
Subsequently, Bonetti et al. \cite{2017PhLB..768..326B} used the confirmed redshift measurement of FRB 121102
to obtain a similar result of $m_{\gamma}\leq3.9\times10^{-50}~\mathrm{kg}$. After correcting for dispersive delay,
Hessels et al. \cite{2019ApJ...876L..23H} found that the subbursts of FRB 121102 still show a time-frequency
downward drifting pattern. The frequency-dependent time delay between subbursts is much smaller than
the dispersive delay, resulting in a tighter upper limit on the photon mass of $m_{\gamma}\leq5.1\times10^{-51}~\mathrm{kg}$
\cite{2019ApJ...882L..13X}. The current astrophysical constraints on the photon rest mass derived through
the dispersion method are shown in Figure~\ref{fig:mass1}. Most of these results were based on a single source,
in which the observed time delay was assumed to be due to the nonzero photon mass and the dispersion from
the plasma effect (see below) was ignored.

\subsection{Dispersion from the plasma effect}\label{subsec:photonmass2}
Due to the dispersive nature of plasma, radio waves with lower frequencies would travel through the ionized median
slower than those with higher frequencies \cite{2017AdSpR..59..736B}. That is, the group velocity of
electromagnetic waves propagating through a plasma has a frequency dependence, i.e.,
\begin{equation}
\upsilon_{p}=c\left[1-\left(\frac{\nu_{p}}{\nu}\right)^{2}\right]^{1/2}\;,
\end{equation}
where the plasma frequency $\nu_{p}=[n_{e}e^{2}/(4\pi^{2} m_{e}\epsilon_{0})]^{1/2}$ with $n_{e}$ the average
electron number density along the line of sight, $e$ and $m_{e}$ the charge and mass of an electron, respectively,
and $\epsilon_{0}$ the permittivity of vacuum. The arrival time delay between two wave packets with different
frequencies, which caused by the plasma effect, can then be expressed as
\begin{equation}\label{eq:tDM}
\begin{aligned}
  \Delta{t_{\rm DM}}&=\int \frac{{\rm d}l}{c} \frac{\nu^{2}_{p}}{2}\left(\nu_l^{-2}-\nu_h^{-2}\right)\\
  &=\frac{e^{2}}{8\pi^{2} m_{e}\epsilon_{0}c}\left(\nu_l^{-2}-\nu_h^{-2}\right){\rm DM_{astro}}\;.
\end{aligned}
\end{equation}
Here the dispersion measure (DM) is defined as the integrated electron number density along the propagation path,
${\rm DM_{astro}}\equiv\int n_{e}{\rm d}l$. In a cosmological context, the measured $\rm DM_{astro}$ by an earth
observer is ${\rm DM_{astro}}\equiv\int n_{e,z}(1+z)^{-1}{\rm d}l$, where $n_{e,z}$ is the rest-frame number density
of free electrons \cite{2014ApJ...783L..35D}. For a cosmological source at redshift $z$, we expect $\rm DM_{astro}$
to separate into four components:
\begin{equation}\label{eq:DMastro}
  {\rm DM_{astro}=DM_{MW}+DM_{MWhalo}+DM_{IGM}+}\frac{\rm DM_{host}}{1+z}
\end{equation}
with $\rm DM_{MW}$ the contribution from the Milky Way ionized interstellar medium, $\rm DM_{MWhalo}$ the contribution from
the Milky Way halo, $\rm DM_{IGM}$ the contribution from the intergalactic medium (IGM), and $\rm DM_{host}$ the contribution
from the host galaxy and source environment in the cosmological rest frame of the source. The measured value of $\rm DM_{host}$
is smaller by a factor of $(1+z)$ \cite{2014ApJ...783L..35D}.

It is evident from Eq.~(\ref{eq:tDM}) that radio waves propagating through a plasma are expected to arrive with
a frequency-dependent dispersion in time of the $1/\nu^{2}$ behavior. However, a similar
dispersion $\propto m_\gamma^2/\nu^2$ [see Eq.~(\ref{eq:tmr})] could also be caused by a nonzero photon mass.
The dispersion method used for constraining the photon mass is, therefore, hindered by the similar frequency dependences
of the dispersions arise from the plasma and nonzero photon mass effects.

\subsection{Combined limits on the photon mass}\label{subsec:photonmass3}
In order to diagnose an effect as radical as a finite photon mass, statistical and possible systematic uncertainties
must be carefully handled. One can not rely solely on a single source, for which it would be hard
to distinguish the dispersions from the plasma and photon mass. For this reason, Shao \& Zhang \cite{2017PhRvD..95l3010S}
constructed a Bayesian framework to derive a combined constraint of $m_\gamma\leq8.7\times10^{-51}~\mathrm{kg}$
from a sample of 21 FRBs (including 20 FRBs without measured redshift, and one, FRB 121102, with a known redshift),
where an uninformative prior was used for the unknown redshift. Wei \& Wu \cite{2018JCAP...07..045W}
also developed a statistical approach to study this problem by analyzing a catalog of radio sources with measured DMs.
This technique has the advantage that it can both give a combined limit of the photon mass and estimate an average
$\rm DM_{astro}$ contributed by the plasma effect. Using the measured DMs from two statistical samples of extragalactic
radio pulsars, Wei \& Wu \cite{2018JCAP...07..045W} placed combined limits on the photon mass at 68\% confidence level,
i.e., $m_\gamma\leq1.5\times10^{-48}~\mathrm{kg}$ for the sample of 22 LMC pulsars and
$m_\gamma\leq1.6\times10^{-48}~\mathrm{kg}$ for the other sample of 5 SMC pulsars.

Since the dispersions from the plasma and photon mass have different redshift dependences,
Refs. \cite{2016PhLB..757..548B,2017PhLB..768..326B,2017AdSpR..59..736B} suggested that they
could in principle be distinguished by a statistical sample of FRBs at a range of different redshifts,
thereby improving the sensitivity to $m_{\gamma}$. Recently, nine FRBs with different redshift measurements
(FRB 121102: $z=0.19723$ \cite{2016Natur.531..202S,2017Natur.541...58C,2017ApJ...834L...7T};
FRB 180916.J0158+65: $z=0.0337$ \cite{2020Natur.577..190M}; FRB 180924: $z=0.3214$ \cite{2019Sci...365..565B};
FRB 181112: $z=0.4755$ \cite{2019Sci...366..231P}; FRB 190523: $z=0.66$ \cite{2019Natur.572..352R};
FRB 190102: $z=0.291$; FRB 190608 $z=0.1178$; FRB 190611: $z=0.378$; FRB 190711: $z=0.522$ \citep{2020Natur.581..391M})
have been reported. Wei \& Wu \cite{2020arXiv200609680W} applied the idea suggested by Refs.
\cite{2016PhLB..757..548B,2017PhLB..768..326B,2017AdSpR..59..736B} to these nine localized FRBs to give a
combined constraint on the photon mass. From observations, the radio signals of all FRBs exhibit an apparent
$\nu^{-2}$-dependent time delay, which is expected from both the free electron content along the line of the sight
and nonzero mass effects on photon propagation (see Eqs.~\ref{eq:tmr} and \ref{eq:tDM}).
Thus, Wei \& Wu \cite{2020arXiv200609680W} attributed the observed time delay to two terms:
\begin{equation}
\Delta{t_{\rm obs}}=\Delta{t_{\rm DM}}+\Delta{t_{m_{\gamma}}}\;.
\end{equation}
In practice, the observed DM of each FRB, $\rm DM_{obs}$, is directly determined by fitting
the $\nu^{-2}$ behavior of its observed time delay. This implies that both the line-of-sight free electron
density and a massive photon (if it exists) determine the same $\rm DM_{obs}$, i.e.,
\begin{equation}\label{eq:DMobs}
{\rm DM_{obs}}={\rm DM_{astro}}+{\rm DM_{\gamma}}\;,
\end{equation}
where $\rm DM_{astro}$ is given in Eq.~(\ref{eq:DMastro}) and $\rm DM_{\gamma}$ denotes the ``effective DM''
due to a nonzero photon mass \cite{2017PhRvD..95l3010S},
\begin{equation}\label{DMr}
{\rm DM_{\gamma}}\equiv\frac{4\pi^2m_e\epsilon_{0}c^5}{h^2e^2}\frac{H_{\gamma}(z)}{H_0}m_\gamma^2\;.
\end{equation}

   \begin{figure}
   \centering
   \vskip-0.1in
   \includegraphics[width=0.5\textwidth, angle=0]{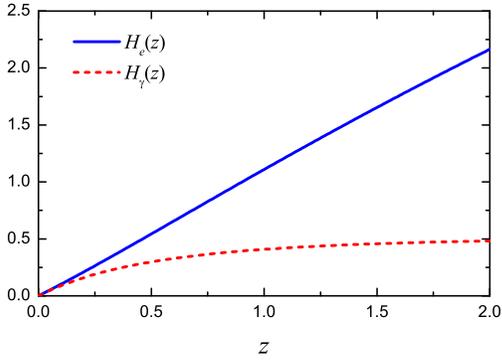}
   \vskip-0.1in
   \caption{Dependence of functions $H_{\gamma}(z)$ and $H_{e}(z)$ on the redshift $z$.
   Reproduced from Ref. \cite{2020arXiv200609680W}.}
   \label{fig:mass2}
   \end{figure}

To investigate the possible $\rm DM_{\gamma}$, we need to figure out different DM contributions in Eq.~(\ref{eq:DMastro}).
For a localized FRB, the $\rm DM_{MW}$ term can be well estimated from a model of our Galactic electron distribution.
The $\rm DM_{MWhalo}$ term is not well modeled, but is expected to contribute about 50--80 pc ${\rm cm^{-3}}$
\cite{2019MNRAS.485..648P}. Wei \& Wu \cite{2020arXiv200609680W} adopted $\rm DM_{MWhalo}=50$ pc ${\rm cm^{-3}}$.
The $\rm DM_{host}$ term is highly uncertain, due to the dependences on the type of the FRB host galaxy,
the relative orientations of the galaxy disk and source, and the near-source plasma \citep{2015RAA....15.1629X}.
Wei \& Wu \cite{2020arXiv200609680W} assumed that the rest-frame $\rm DM_{host}$ scales with the star formation rate
(SFR; see Ref. \cite{2018MNRAS.481.2320L} for more details),
${\rm DM_{host}}(z)={\rm DM_{host,0}}\sqrt{{\rm SFR}(z)/{\rm SFR}(0)}$, where $\rm DM_{host,0}$
represents the present value of ${\rm DM_{host}}(z=0)$ and
${\rm SFR}(z)=\frac{0.0157+0.118z}{1+(z/3.23)^{4.66}}$ $\rm M_{\odot}$ $\rm yr^{-1}$ ${\rm Mpc^{-3}}$
is the empirical form of the star formation history \cite{2006ApJ...651..142H,2008MNRAS.388.1487L}.
In the analysis of Wei \& Wu \cite{2020arXiv200609680W}, $\rm DM_{host,0}$ was treated as a free parameter.
The IGM contribution to the DM of an FRB at redshift $z$ is related to the ionization fractions of
hydrogen and helium in the universe. Since both hydrogen and helium are fully ionized at $z<3$, one then has
\citep{2014ApJ...783L..35D}
\begin{equation}\label{eq:DMIGM}
{\rm DM_{IGM}}(z)=\frac{21cH_{0}\Omega_{b}f_{\rm IGM}}{64\pi G m_p}H_{e}(z)\;,
\end{equation}
where $m_p$ is the proton mass, $\Omega_{b}=0.0493$ is the baryonic matter energy density \cite{2020A&A...641A...6P},
$f_{\rm IGM}\simeq0.83$ is the fraction of baryon mass in the IGM \cite{1998ApJ...503..518F},
and $H_{e}(z)$ is the dimensionless redshift function,
\begin{equation}\label{eq:He}
  H_{e}(z)=\int_{0}^{z}\frac{(1+z')dz'}{\sqrt{\Omega_{\rm m}(1+z')^{3}+\Omega_{\Lambda}}}\;.
\end{equation}
The two dimensionless quantities ($H_{e}$ and $H_{\gamma}$) as a function of the redshift $z$
are displayed in Figure~\ref{fig:mass2}. One can see from this plot that the IGM and a possible
photon mass contributions to DMs have different redshift dependences. As already commented, Bonetti et al.
\cite{2016PhLB..757..548B,2017PhLB..768..326B} indicated that the different redshift dependences
might not only be able to break dispersion degeneracy but also improve the sensitivity to $m_{\gamma}$
at the point when a few redshift measurements of FRBs become available (see also Refs.
\cite{2017AdSpR..59..736B,2017PhRvD..95l3010S}).

   \begin{figure}
   \centering
   \includegraphics[width=0.5\textwidth, angle=0]{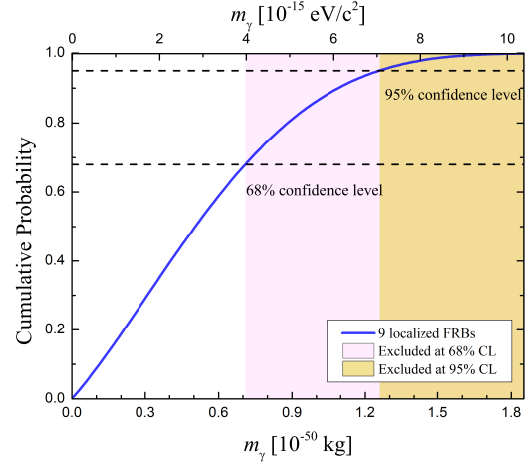}
   \vskip-0.2in
   \caption{Cumulative posterior probability distribution on the photon rest mass
   $m_{\gamma}$ derived from nine localized FRBs. The excluded values for $m_{\gamma}$
   at 68\% and 95\% confidence levels are displayed with shadowed areas.
   Reproduced from Ref. \cite{2020arXiv200609680W}.}
   \label{fig:mass3}
   \end{figure}

\begin{table*}
\caption{Summary of astrophysical upper bounds on the photon rest mass as obtained by the dispersion method.}
\small
\centering
\begin{tabular}{lllcc}
\hline
\hline
              &           & Wavelength (energy or  &                               &      \\
Author (year) & Source(s) & frequency) range       &  $m_{\gamma}~\mathrm{(kg)}$   & Refs. \\
\hline
Lovell et al. (1964) & flare stats & 0.54 $\mu$m--1.2 m & $1.6\times10^{-45}$ & \cite{1964Natur.202..377L} \\
Warner and Nather (1969) & Crab Nebula pulsar  & 0.35--0.55 $\mu$m & $5.2\times10^{-44}$ & \cite{1969Natur.222..157W} \\
Schaefer (1999) & GRB 980703  & $5.0\times10^{9}$--$1.2\times10^{20}$ Hz & $4.2\times10^{-47}$ & \cite{1999PhRvL..82.4964S} \\
Zhang et al. (2016) & GRB 050416A  & 8.46 GHz--15 keV & $1.1\times10^{-47}$ & \cite{2016JHEAp..11...20Z} \\
Wei et al. (2017) & Extragalactic radio pulsar  & $\sim1.4$ GHz & $2.0\times10^{-48}$ & \cite{2017RAA....17...13W} \\
                  & (PSR J0451-67)          &   &   &   \\
                  & Extragalactic radio pulsar  & $\sim1.4$ GHz & $2.3\times10^{-48}$ &  \\
                  & (PSR J0045-7042)          &   &   &   \\
Wei and Wu (2018) & 22 radio pulsars in the LMC  & $\sim1.4$ GHz & $1.5\times10^{-48}$ & \cite{2018JCAP...07..045W} \\
                  & 5 radio pulsars in the SMC   & $\sim1.4$ GHz & $1.6\times10^{-48}$ &  \\
Wu et al. (2016)  & FRB 150418      & 1.2--1.5 GHz & $5.2\times10^{-50}$ & \cite{2016ApJ...822L..15W} \\
Bonetti et al. (2016)  & FRB 150418      & 1.2--1.5 GHz & $3.2\times10^{-50}$ & \cite{2016PhLB..757..548B}\\
Bonetti et al. (2017)  & FRB 121102      & 1.1--1.7 GHz & $3.9\times10^{-50}$ & \cite{2017PhLB..768..326B}\\
Shao and Zhang (2017)  & 21 FRBs (20 of them without  & $\sim$ GHz & $8.7\times10^{-51}$ & \cite{2017PhRvD..95l3010S} \\
                       & redshift measurement)        &            &                     &                            \\
Xing et al. (2019)  & FRB 121102 subbursts      & 1.34--1.37 GHz & $5.1\times10^{-51}$ & \cite{2019ApJ...882L..13X}\\
Wei and Wu (2020) & 9 localized FRBs  & $\sim$ GHz & $7.1\times10^{-51}$ & \cite{2020arXiv200609680W} \\
\hline
\end{tabular}
\label{table3}
\end{table*}

With the redshift measurements of nine FRBs, Wei \& Wu \cite{2020arXiv200609680W} maximized
the likelihood function,
\begin{equation}\label{eq:likelihood}
\begin{aligned}
\mathcal{L} = &\prod_{i}
\frac{1}{\sqrt{2\pi}\sigma_{{\rm tot},i}}\\
&\times\exp\left\{-\,\frac{\left[{{\rm DM}_{{\rm obs}, i}-{\rm DM}_{{\rm astro}, i}-{\rm DM}_{\gamma}}(m_{\gamma},z_i)\right]^{2}}
{2\sigma_{{\rm tot},i}^{2}}\right\}\;,
\end{aligned}
\end{equation}
to derive a combined limit on the photon mass $m_{\gamma}$.
Here the total variance on each FRB is given by
\begin{equation}
\sigma_{\rm tot}^{2}=\sigma_{\rm obs}^{2}+\sigma_{\rm MW}^{2}+\sigma_{\rm MWhalo}^{2}+\sigma_{\rm int}^{2}\;,
\end{equation}
where $\sigma_{\rm obs}$, $\sigma_{\rm MW}$, and $\sigma_{\rm MWhalo}$ correspond to the uncertainties of
$\rm DM_{obs}$, $\rm DM_{MW}$, and $\rm DM_{MWhalo}$, respectively, and $\sigma_{\rm int}$
represents the global intrinsic scatter that might originate from the diversity of host galaxy contribution
and the large IGM fluctuation. The marginalized cumulative posterior distribution of the photon mass $m_{\gamma}$
is shown in Figure~\ref{fig:mass3}. The 68\% confidence-level upper limit on $m_{\gamma}$ from nine localized FRBs
is $m_{\gamma}\leq7.1\times10^{-51}\;{\rm kg}$ \cite{2020arXiv200609680W}, which is comparable with or
represents a factor of 7 improvement over existing photon mass limits from the individual FRBs \cite{2016ApJ...822L..15W,2016PhLB..757..548B,2017PhLB..768..326B,2019ApJ...882L..13X}.

Table~\ref{table3} presents a summary of astrophysical upper bounds on the photon mass $m_{\gamma}$ obtained through
the dispersion method. As illustrated in Figure~\ref{fig:mass1} and Table~\ref{table3}, the current best
$m_{\gamma}$ limits were made by using the observations of FRB 121102 subbursts ($m_{\gamma}\leq5.1\times10^{-51}~\mathrm{kg}$)
\cite{2019ApJ...882L..13X} and nine localized FRBs ($m_{\gamma}\leq7.1\times10^{-51}\;{\rm kg}$) \cite{2020arXiv200609680W}.

\section{
Astrophysical tests of the WEP}\label{sec:WEP}
The WEP states that inertial and gravitational masses are identical. An alternative statement is that
the trajectory of a freely falling, uncharged test body is independent of its internal structure and composition.
Astrophysical observations provide a unique test of WEP by testing if different massless particles experience
gravity differently. In this section, we present the field-standard test method through the Shapiro time delay effect.

\begin{figure}
   \centering
   \includegraphics[width=0.4\textwidth, angle=0]{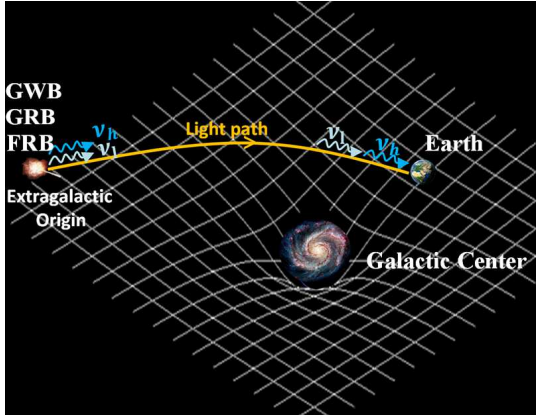}
   \caption{A cartoon picture shows how two photons, one at a low frequency ($\nu_{l}$) and another at a high frequency ($\nu_{h}$), travel in curved space-time from their origin in a distant extragalactic transient until reaching our Earth. A lower-limit estimate of the gravitational pull that the photons experience along their way is given by the mass in the center of the Milky Way.}
   \label{fig:WEP1}
   \end{figure}

Particles which traverse a gravitational field would experience a time delay (named the Shapiro delay)
due to the warping of spacetime. Adopting the PPN formalism, the $\gamma$-dependent Shapiro delay is
given by \cite{1964PhRvL..13..789S}
\begin{equation}
t_{\rm gra}=-\frac{1+\gamma}{c^3}\int_{r_e}^{r_o}~U(r)dr\;,
\label{eq:Shapiro}
\end{equation}
where $r_e$ and $r_o$ denote the locations of the emitting source and observer, respectively, and $U(r)$ is
the gravitational potential along the propagation path. A possible violation of the WEP implies that,
if two particles follow the same path through a gravitational potential, then they would undergo different
Shapiro delays. In this case, two test particles emitted simultaneously from the source would reach our Earth
with a arrival-time difference (see Figure~\ref{fig:WEP1})
\begin{equation}
\Delta t_{\rm gra}=\frac{\Delta\gamma}{c^3}\int_{r_e}^{r_o}~U(r)dr\;,
\label{eq:delta-tgra}
\end{equation}
where $\Delta\gamma=|\gamma_{2}-\gamma_{1}|$ represents the difference of the $\gamma$ values for different
particles, which can be used as measure of a possible deviation from the WEP.

To estimate the relative Shapiro delay $\Delta t_{\rm gra}$ with Eq.~(\ref{eq:delta-tgra}), one needs to
figure out the gravitational potential $U(r)$. For sources at cosmological distances, $U(r)$ should have
contributions from the local gravitational potential $U_{\rm local}(r)$, the intergalactic potential $U_{\rm IG}(r)$,
and the host galaxy potential $U_{\rm host}(r)$. Since the potential function for $U_{\rm IG}(r)$ and
$U_{\rm host}(r)$ is hard to model, for the purposes of obtaining lower limits, it is reasonable to extend
the local potential $U_{\rm local}(r)$ to the distance of the source. In previous articles, the gravitational
potential of the Milky Way, Virgo Cluster, or the Laniakea supercluster \cite{2014Natur.513...71T} has been
adopted as the local potential, which can be modeled as a Keplerian potential $U(r)=-GM/r$.
We thus have \cite{1988PhRvL..60..173L,2016PhRvD..94b4061W}
\begin{eqnarray}
\Delta t_{\rm gra}= \Delta\gamma \frac{GM}{c^{3}} \times \qquad\qquad\qquad\qquad\qquad\qquad\qquad\\ \nonumber
\ln \left\{ \frac{ \left[d+\left(d^{2}-b^{2}\right)^{1/2}\right] \left[r_{L}+s_{\rm n}\left(r_{L}^{2}-b^{2}\right)^{1/2}\right] }{b^{2}} \right\}\;,
\label{eq:gammadiff}
\end{eqnarray}
where $M$ is the mass of the gravitational field source, $d$ is the distance from the particle source to the center of
gravitational field, $b$ is the impact parameter of the particle paths relative to the center,
$r_{L}$ is the distance of the center, and $s_{\rm n}=+1$ (or $-1$) corresponds to the case where
the particle source is located in the same (or opposite) direction with respect to the center of gravitational field.
For a particle source at the coordinates (R.A.=$\beta_{s}$, Dec.=$\delta_{s}$), the impact parameter $b$
can be estimated as
\begin{equation}
b=r_{G}\sqrt{1-(\sin \delta_{s} \sin \delta_{L}+\cos \delta_{s} \cos \delta_{L} \cos(\beta_{s}-\beta_{L}))^{2}}\;,
\end{equation}
where $\beta_{L}$ and $\delta_{L}$ represent the coordinates of the center of gravitational field.

\subsection{Arrival time tests of the WEP}\label{subsec:WEP1}
If one assumes that the observed time delay ($\Delta t_{\rm obs}$) between messengers or within messengers
is mainly contributed by the relative Shapiro delay ($\Delta t_{\rm gra}$) and we know that the intrinsic
(astrophysical) time delay $\Delta t_{\rm int}>0$, a conservative upper limit on the WEP violation
could be placed by\footnote{Minazzoli et al. \cite{2019PhRvD.100j4047M} discussed the shortcomings of standard
Shapiro delay-based tests of the WEP. Because such tests are based on the estimation for different messenger particles
of the one-way propagation time between the emitting source and the observer that is not an observable per se.
As a consequence, Minazzoli et al. \cite{2019PhRvD.100j4047M} suggested that such tests are extremely model dependent
and can not provide reliable quantitative limits on the WEP violation.}
\begin{eqnarray}
\Delta\gamma < \Delta t_{\rm obs} \left(\frac{GM}{c^{3}}\right)^{-1} \times \qquad\qquad\qquad\qquad\qquad\qquad\qquad\\ \nonumber
\ln^{-1} \left\{ \frac{ \left[d+\left(d^{2}-b^{2}\right)^{1/2}\right] \left[r_{L}+s_{\rm n}\left(r_{L}^{2}-b^{2}\right)^{1/2}\right] }{b^{2}} \right\}\;.
\end{eqnarray}

The first multimessenger test of the WEP was between photons and neutrinos from supernova SN 1987A.
Logo \cite{1988PhRvL..60..176K} and Krauss \& Tremaine \cite{1988PhRvL..60..173L} used the observed time delay
between photons and MeV neutrinos from SN 1987A to prove that the $\gamma$ values for photons and neutrinos are
identical to an accuracy of approximately 0.3--0.5\%. These constraints can be further improved using the likely
associations of high-energy neutrinos with the flaring blazars \cite{2016PhRvL.116o1101W,2019EPJC...79..185B,
2019PhRvD.100j3002L,2019JHEAp..22....1W} or GRBs \cite{2016JCAP...08..031W}. Besides the neutrino-photon sectors,
the coincident detections of GW events with electromagnetic counterparts also provided multimessenger tests of
the WEP, extending the WEP tests with GWs and photons \cite{2016PhRvD..94b4061W,2016ApJ...827...75L,2017ApJ...848L..13A,
2017PhLB..770....8L,2017ApJ...851L..18W,2017JCAP...11..035W,2018PhRvD..97h3013S,2018PhRvD..97d1501B,2020ApJ...900...31Y}.
For example, with the assumption that the arrival time delay between GW170817 and GRB 170817A ($\sim1.7$ s) from a binary
neutron star merger is mainly due to the gravitational potential of the Milky Way outside a sphere of 100 kpc,
Abbott et al. \cite{2017ApJ...848L..13A} derived $-2.6\times10^{-7}\leq \gamma_{g}-\gamma_{\gamma}\leq1.2\times10^{-6}$.
A more severe constraint of $|\gamma_{g}-\gamma_{\gamma}|<0.9\times10^{-10}$ can be achieved for GW170817/GRB 170817A
when the gravitational potential of the Virgo Cluster is considered \cite{2017JCAP...11..035W}.

WEP tests have also been performed within the same species of messenger particles (neutrinos, photons, or GWs)
but with varying energies (e.g., \cite{1988PhRvL..60..173L,2015ApJ...810..121G,2015PhRvL.115z1101W,2016PhRvD..94b4061W,1999BASI...27..627S,
2016MNRAS.460.2282S,2016JHEAp...9...35L,2018ApJ...860..173Y,2016ApJ...820L..31T,2016ApJ...821L...2N,
2019ApJ...882L..13X,2020PDU....2900571W,2016ApJ...818L...2W,2016PhRvD..94j1501Y,2017ApJ...837..134Z,
2018EPJC...78...86D,2018ApJ...861...66L,2016PhLB..756..265K,2020MNRAS.499L..53Y}).
In the neutrino sector, Longo \cite{1988PhRvL..60..173L} used the observed delay between 7.5 MeV and 40 MeV neutrinos
from SN 1987A to set $|\gamma_{\nu}({\rm 40\;MeV})-\gamma_{\nu}({\rm 7.5\;MeV})|<1.6\times10^{-6}$.
Wei et al. \cite{2019JHEAp..22....1W} adopted the delay for neutrinos ranging in energy from about 0.1 to 20 Tev
from the direction of the blazar TXS 0506+056 to obtain $|\gamma_{\nu}({\rm 20\;TeV})-\gamma_{\nu}({\rm 0.1\;TeV})|<7.3\times10^{-6}$. In the photon sector, Gao et al. \cite{2015ApJ...810..121G} (see also
Sivaram \cite{1999BASI...27..627S}) suggested that one can use the time delays of photons of different energies
from cosmological transients to test the WEP. They applied the similar approach to GRBs and derived
$|\gamma_{\gamma}({\rm eV})-\gamma_{\gamma}({\rm MeV})|<1.2\times10^{-7}$ for GRB 080319B and
$|\gamma_{\gamma}({\rm GeV})-\gamma_{\gamma}({\rm MeV})|<2.0\times10^{-8}$ for GRB 090510.
Recently, such a test has also been applied to different-energy photons in other transient sources,
including FRBs \cite{2015PhRvL.115z1101W,2016ApJ...820L..31T,2016ApJ...821L...2N,2019ApJ...882L..13X,2020PDU....2900571W},
TeV blazars \cite{2016ApJ...818L...2W}, and the Crab pulsar \cite{2016PhRvD..94j1501Y,2017ApJ...837..134Z,2018EPJC...78...86D,2018ApJ...861...66L}.
In the GW sector, Wu et al. \cite{2016PhRvD..94b4061W} suggested that one can treat the GW signals with different
frequencies as different gravitons to test the WEP. They used the delay for the GW signals ranging in frequency
from about 35 to 150 Hz from GW150914 to set $|\gamma_{g}({\rm 35\;Hz})-\gamma_{g}({\rm 150\;Hz})|<10^{-9}$
(see also \cite{2016PhLB..756..265K}). Very recently, Yang et al. \cite{2020MNRAS.499L..53Y} obtained new constraints of
$\Delta \gamma$ by using the GW data of binary black holes mergers in the LIGO-Virgo catalogue GWTC-1.
The best constraints came from GW170104 and GW170823, i.e., $\Delta \gamma<10^{-15}$.

Yu \& Wang \cite{2018EPJC...78..692Y} and Minazzoli \cite{2019arXiv191206891M} proposed a new multimessenger test
of the WEP using strongly lensed cosmic transients. By measuring the time delays between lensed images seen in
different messengers, one can obtain robust constraints on the differences of the $\gamma$ values.

\subsection{Polarization tests of the WEP}\label{subsec:WEP2}
The fact that the trajectory of any freely falling test body does not depend on its internal structure
is one of the consequences of the WEP. Since the polarization is viewed as a basic component of the
internal structure of photons, Wu et al. \cite{2017PhRvD..95j3004W} proposed that multiband electromagnetic
emissions exploiting different polarizations are an essential tool for testing the accuracy of the WEP.

For a linearly polarized light, it is the combination of two monochromatic waves with opposite circular
polarizations (labeled with `$l$' and `$r$'). If the WEP fails, then the two circularly polarized beams
travel through the same gravitational field with different Shapiro delays. The relative Shapiro delay
$\Delta t_{\rm gra}$ of these two beams is the same as Eq.~(\ref{eq:delta-tgra}), except now
$\Delta\gamma=|\gamma_{l}-\gamma_{r}|$ corresponds to the difference of the $\gamma$ values for
the left- and right-handed circular polarization states. Because of the arrival-time difference between
the two circular components, the polarization plane of a linearly polarized light would be subject
to a rotation along the photons' propagation path. The rotation angle due to the WEP violation is given by
\cite{2017MNRAS.469L..36Y,2019PhRvD..99j3012W}
\begin{equation}
\Delta\phi_{\rm WEP}\left(E\right)=\Delta t_{\rm gra}\frac{2\pi c}{\lambda}=\Delta t_{\rm gra}\frac{E}{\hbar}\;,
\label{eq:theta-WEP}
\end{equation}
where $E$ is the observed energy.

\begin{figure*}
\centering
\vskip-0.2in
\includegraphics[width=0.8\textwidth]{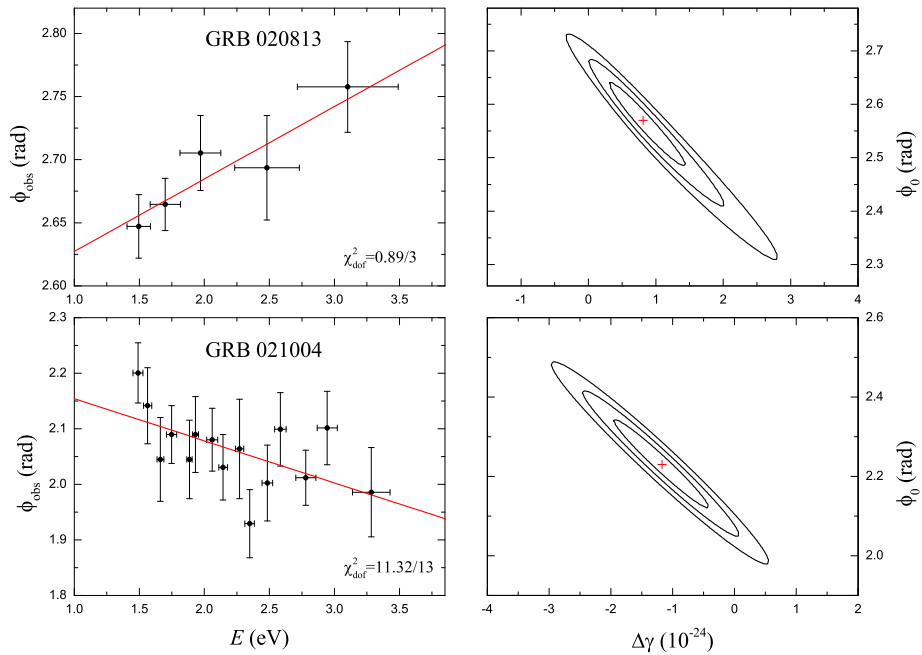}
\vskip-0.3in
\caption{Fit to the afterglow polarimetry data of GRB 020813 and GRB 021004.
Left panels: linear fits of the spectropolarimetric data. Right panels: 1-3$\sigma$ confidence levels
in the $\phi_{0}$-$\Delta \gamma$ plane. Reproduced from Ref. \cite{2020EPJP..135..527W}.}
\label{fig:WEP2}
\end{figure*}

Since the initial angle of the linearly polarized light is not available, the exact value of $\Delta\phi_{\rm WEP}$
is unknown. Yet, we can set an upper limit for the differential rotation angle $|\Delta\phi(E_{2})-\Delta\phi(E_{1})|$,
because the polarization signal would be erased as the path difference goes beyond the coherence length.
Therefore, the detection of liner polarization indicates that $|\Delta\phi(E_{2})-\Delta\phi(E_{1})|$ should
not be too large. By considering the Shapiro delay of the Milky Way's gravitational potential, and setting
the upper limit of $|\Delta\phi(E_{2})-\Delta\phi(E_{1})|$ to be $2\pi$, Yang et al. \cite{2017MNRAS.469L..36Y}
obtained an upper limit on the $\gamma$ discrepancy of $\Delta \gamma<1.6\times10^{-27}$ from the polarization
measurement of GRB 110721A.\footnote{Similar LIV constraints have been obtained by astrophysical polarization
measurements \cite{2007MNRAS.376.1857F,2012PhRvL.109x1104T}. Here, the LIV effect was supposed not to work
simultaneously to accidentally enhance or cancel the effect from the WEP violation.}
Through a detailed calculation on the evolution of GRB polarization arising from the WEP violation,
Wei \& Wu \cite{2019PhRvD..99j3012W} proved that the initial polarization signal is not significantly suppressed
even if $|\Delta\phi(E_{2})-\Delta\phi(E_{1})|$ is larger than $\pi/2$. Applying their formulae to the GRB
polarimetric data, Wei \& Wu \cite{2019PhRvD..99j3012W} placed the most stringent limits so far on a deviation
from the WEP for two cases: $\Delta \gamma<0.8\times10^{-33}$ for GRB 061122 and $\Delta \gamma<1.3\times10^{-33}$
for GRB 110721A.

If the rotation angle induced by the WEP violation is considered here, the observed linear polarization angle
for photons emitted at a certain energy $E$ with an intrinsic polarization angle $\phi_{0}$ should be
\begin{equation}
\phi_{\rm obs}=\phi_{0}+\Delta\phi_{\rm WEP}\left(E\right)\;.
\end{equation}
Instead of requiring the argument that the rotation of the polarization plane would severely reduce polarization
over a broad bandwidth, Wei \& Wu \cite{2020EPJP..135..527W} simply assumed that $\phi_{0}$ is an unknown constant.
One can then constrain the WEP violation by measuring the energy-dependent change of the polarization angle.
Wei \& Wu \cite{2020EPJP..135..527W} showed that it is possible to give constraints on both $\Delta \gamma$
and $\phi_{0}$ through direct fitting of the multiwavelength polarimetric data of the optical afterglows of GRB 020813
and GRB 021004 (see Figure~\ref{fig:WEP2}). At the $3\sigma$ confidence level, the joint constraint on $\Delta \gamma$
from two GRBs is $-2.7\times10^{-24}<\Delta \gamma<3.1\times10^{-25}$.
Yi et al. \cite{2020MNRAS.493.1782Y,2020MNRAS.498.4295Y} applied the same method to radio polarimetry data of
the blazar 3C 279 and extragalactic radio sources, and obtained less stringent constraints on $\Delta \gamma$.

Shapiro delay-based tests of the WEP are summarized in Table~\ref{table4} for comparison. When the test particles
are different messengers, the best multimessenger constraint is $\Delta \gamma<1.3\times10^{-13}$ for keV photons
and the TeV neutrino from GRB 110521B \cite{2016JCAP...08..031W}. When the test particles are the same messengers
but with different energies, the best constraints are $\Delta \gamma<2.5\times10^{-16}$ for 1.344--1.374 GHz photons
from FRB 121102 subbursts \cite{2019ApJ...882L..13X} and $\Delta \gamma<6.2\times10^{-16}$ for $\sim100$ Hz GW signals
from GW170104 \cite{2020MNRAS.499L..53Y}. When the test particles are the same messengers but with different polarization
states, the best constraint is $\Delta \gamma<0.8\times10^{-33}$ for polarized gamma-ray photons from GRB 061122
\cite{2019PhRvD..99j3012W}.

\begin{sidewaystable*}[h]
\tiny
\caption{Upper constraints on the differences of the $\gamma$ values through the Shapiro (gravitational) time delay effect.}
\begin{tabular}{llllccc}
&&&&& \\
\hline\hline
&&&&& \\
{Categorization}&{Author (year)}&{Source}&{Messengers}&{Gravitational Field}&{$\Delta\gamma$}&{Refs.}\\
&&&&& \\
\hline
    & Longo (1988) & SN 1987A & eV photons and MeV neutrinos & Milky Way & $3.4\times10^{-3}$ & \cite{1988PhRvL..60..173L}\\
    & Krauss and Tremaine (1988) & SN 1987A & eV photons and MeV neutrinos & Milky Way & $5.0\times10^{-3}$ & \cite{1988PhRvL..60..176K}\\
    & Wang et al. (2016)  & PKS B1424-418 & MeV photons and PeV neutrino & Virgo Cluster & $3.4\times10^{-4}$ & \cite{2016PhRvL.116o1101W}\\
    &                     & PKS B1424-418 & MeV photons and PeV neutrino & Great Attractor & $7.0\times10^{-6}$ & \\
    & Boran et al. (2019) & TXS 0506+056 & GeV photons and TeV neutrino & Milky Way & $5.5\times10^{-2}$ & \cite{2019EPJC...79..185B}\\
    & Laha (2019); Wei et al. (2019)  & TXS 0506+056 & GeV photons and TeV neutrino & Laniakea supercluster of galaxies & $10^{-6}$--$10^{-7}$ & \cite{2019PhRvD.100j3002L,2019JHEAp..22....1W}\\
~Different messengers    & Wei et al. (2016a)  & GRB 110521B & keV photons and TeV neutrino  & Laniakea supercluster of galaxies & $1.3\times10^{-13}$ & \cite{2016JCAP...08..031W}\\
    & Abbott et al. (2017)  & GW170817 & MeV photons and GW signals  & Milky Way   & -2.6$\times10^{-7}$---1.2$\times10^{-6}$ & \cite{2017ApJ...848L..13A}\\
    & Wei et al. (2017)  & GW170817 & MeV photons and GW signals  & Virgo Cluster  & 9.2$\times10^{-11}$ & \cite{2017JCAP...11..035W}\\
    &                             & GW170817 & eV photons and GW signals   & Virgo Cluster  & 2.1$\times10^{-6}$ & \\
    & Shoemaker and Murase (2018)  & GW170817 & MeV photons and GW signals  & Milky Way  & 7.4$\times10^{-8}$ & \cite{2018PhRvD..97h3013S}\\
    & Boran et al. (2018)  & GW170817 & MeV photons and GW signals  & Milky Way  & 9.8$\times10^{-8}$ & \cite{2018PhRvD..97d1501B}\\
    & Yao et al. (2020)  & GW170817 & MeV photons and GW signals  & Milky Way $+$ host galaxy  & $\sim10^{-9}$ & \cite{2020ApJ...900...31Y}\\
    &                    & GW170817 & eV photons and GW signals   & Milky Way $+$ host galaxy  & $\sim10^{-4}$  & \\
\hline
    & Longo (1988) & SN 1987A & 7.5--40 MeV neutrinos & Milky Way & $1.6\times10^{-6}$ & \cite{1988PhRvL..60..173L} \\
    & Wei et al. (2019)  & TXS 0506+056 & 0.1--20 TeV neutrinos & Laniakea supercluster of galaxies & $7.3\times10^{-6}$ & \cite{2019JHEAp..22....1W}\\
    & Sivaram (1999)  & GRB 990123 & eV--MeV photons & Milky Way & $4.0\times10^{-7}$ & \cite{1999BASI...27..627S} \\
    & Gao et al. (2015)  & GRB 090510 & MeV--GeV photons & Milky Way & $2.0\times10^{-8}$ & \cite{2015ApJ...810..121G} \\
    &                    & GRB 080319B & eV--MeV photons & Milky Way & $1.2\times10^{-7}$ & \\
    & Wei et al. (2015)  & FRB 110220 & 1.2--1.5 GHz photons & Milky Way & $2.5\times10^{-8}$ & \cite{2015PhRvL.115z1101W}\\
    &                    & FRB/GRB 100704A & 1.23--1.45 GHz photons & Milky Way & $4.4\times10^{-9}$ & \\
~Same messengers    & Tingay and Kaplan (2016) & FRB 150418 & 1.2--1.5 GHz photons & Milky Way & (1--2)$\times10^{-9}$ & \cite{2016ApJ...820L..31T} \\
~~~~~~~~with    & Nusser (2016) & FRB 150418 & 1.2--1.5 GHz photons & Large-scale structure & $10^{-12}$--$10^{-13}$ & \cite{2016ApJ...821L...2N} \\
~different energies    & Xing et al. (2019)         & FRB 121102 subbursts & 1.344--1.374 GHz photons & Laniakea supercluster of galaxies & $2.5\times10^{-16}$ & \cite{2019ApJ...882L..13X}\\
    & Wei et al. (2016b)  & Mrk 421 & keV--TeV photons & Milky Way & $3.9\times10^{-3}$ & \cite{2016ApJ...818L...2W} \\
    &                             & PKS 2155-304 & sub TeV--TeV photons & Milky Way & $2.2\times10^{-6}$ & \\
    & Yang and Zhang (2016)  & Crab pulsar & 8.15--10.35 GHz photons  & Milky Way  & (0.6--1.8)$\times10^{-15}$ & \cite{2016PhRvD..94j1501Y}\\
    & Zhang and Gong (2017)   & Crab pulsar & eV--MeV photons  & Milky Way  & 3.0$\times10^{-10}$ & \cite{2017ApJ...837..134Z} \\
    & Desai and Kahya (2018)  & Crab pulsar & 8.15--10.35 GHz photons  & Milky Way  & 2.4$\times10^{-15}$ & \cite{2018EPJC...78...86D}\\
    & Leung et al. (2018)  & Crab pulsar & 1.52--2.12 eV photons  & Milky Way  & 1.1$\times10^{-10}$ & \cite{2018ApJ...861...66L} \\
    & Wu et al. (2016)  & GW150914 & 35--150 Hz GW signals  & Milky Way  & $\sim10^{-9}$ & \cite{2016PhRvD..94b4061W} \\
    & Kahya and Desai (2016)  & GW150914 & 35--250 Hz GW signals  & Milky Way  & 2.6$\times10^{-9}$ & \cite{2016PhLB..756..265K} \\
    & Yang et al. (2020)  & GW170104 & $\sim100$ Hz GW signals  & Milky Way  & 6.2$\times10^{-16}$ & \cite{2020MNRAS.499L..53Y} \\
    &                         & GW170823 & $\sim100$ Hz GW signals  & Milky Way  & 1.0$\times10^{-15}$ &  \\
\hline
   & Wu et al. (2017) & FRB 150807 & Polarized radio photons  & Laniakea supercluster of galaxies  & $2.2\times10^{-16}$ & \cite{2017PhRvD..95j3004W} \\
   &                  & GRB 120308A & Polarized optical photons  & Laniakea supercluster of galaxies  & $1.2\times10^{-10}$ & \\
~Same messengers   &                  & GRB 100826A & Polarized gamma-ray photons  & Laniakea supercluster of galaxies  & $1.2\times10^{-10}$ & \\
~~with different   & Yang et al. (2017)  & GRB 110721A & Polarized gamma-ray photons  & Milky Way  & $1.6\times10^{-27}$ & \cite{2017MNRAS.469L..36Y}\\
polarization states   & Wei and Wu (2019)  & GRB 061122  & Polarized gamma-ray photons  & Laniakea supercluster of galaxies  & $0.8\times10^{-33}$ & \cite{2019PhRvD..99j3012W} \\
   &                             & GRB 110721A & Polarized gamma-ray photons  & Laniakea supercluster of galaxies  & $1.3\times10^{-33}$ \\
   & Wei and Wu (2020)  & GRB 020813 and  & Polarized optical photons  & Milky Way & -2.7$\times10^{-24}$---3.1$\times10^{-25}$  & \cite{2020EPJP..135..527W} \\
   &                    & GRB 021004  &    &   &    & \\
\hline\hline
\end{tabular}
\label{table4}
\end{sidewaystable*}

\section{
Summary and future prospect}\label{sec:summary}

In this paper, we have reviewed the status of the high-precision tests of fundamental physics with
astrophysical transients. A few solid conclusions and future prospect can be summarized:

(i) Violations of Lorentz invariance can be tested by seeking a frequency-dependent velocity resulting from vacuum dispersion
and by searching for a change in polarization arising from vacuum birefringence.

For vacuum dispersion studies, in order to obtain tighter limits on the LIV effects one should choose the waves
of higher frequencies that propagate over longer distances. GRBs, with their short spectral lags, cosmological distances, and
gamma-ray emissions, are the most powerful probes so far for LIV constraints in the dispersive photon sector.
In the past, emission from GRBs has been observed only at energies below 100 GeV. Recently, VHE photons above 100 GeV
have been detected from GRB 190114C \cite{2019Natur.575..455M} and GRB 180720B \cite{2019Natur.575..464A}, opening a new
window to study GRBs in sub-TeV gamma-rays. Such detections are expected to become routine in the future
\cite{2019Natur.575..448Z}, especially with the operations of facilities such as the international Cherenkov Telescope Array
(CTA) and the Large High Altitude Air Shower Observatory (LHAASO) in China. Observations of extremely high-energy emission
from more GRBs will further improve constraints on LIV. For vacuum birefringence studies, in order to tightly constrain
the LIV effects one should choose those astrophysical sources with larger distances and polarization observations
in a higher energy band. Thanks to their polarized gamma-ray emissions and large cosmological distances, GRBs are
promising sources for seeking LIV-induced vacuum birefringence. As more and more GRB polarimeters (such as POLAR-II,
TSUBAME, NCT/COSI, GRAPE; see McConnell \cite{2017NewAR..76....1M} for a review) enter service, it is reasonable to
expect that the GRB polarimetric data will be significantly enlarged. Limits on the vacuum birefringent effect
can be further improved with larger number of GRBs with higher energy polarimetry and higher redshifts.

(ii) The most direct and robust method for constraining the photon rest mass is to measure the frequency-dependent dispersion
of light.

Theoretically speaking, to obtain more stringent bounds on the photon mass one should choose the waves of lower frequencies
that propagate over longer distances. Cosmological FRBs, with their short time durations, low frequency emissions, and
long propagation distances, are the most excellent astrophysical probes so far for constraining the photon mass. However,
the derivation of a photon mass limit through the dispersion method is complicated by the similar frequency dependences
of the dispersions due to the plasma and nonzero photon mass effects. If in the future more FRB redshifts are measured,
the different redshift dependences of the plasma and photon mass contributions to the DM can be used to break dispersion
degeneracy and to improve the sensitivity to the photon mass
\cite{2016PhLB..757..548B,2017PhLB..768..326B,2017AdSpR..59..736B,2017PhRvD..95l3010S,2020arXiv200609680W}.
It is encouraging that radio telescopes, such as the Canadian Hydrogen Intensity Mapping Experiment (CHIME)
\cite{2020Natur.577..190M}, the Australia Square Kilometer Array Pathfinder (ASKAP) \cite{2019Sci...365..565B},
and the Deep Synoptic Array 10-dish prototype (DSA-10) \cite{2019Natur.572..352R}, have led to direct localizations of FRBs.
The field will also greatly benefit from the operations of facilities, such as the Five Hundred Meter Aperture Spherical
radio Telescope (FAST), the DSA-2000, and the Square Kilometer Array (SKA). The rapid progress in localizing FRBs
will further improve the constraints on the photon mass. In the future, a swarm of nano-satellites orbiting the Moon
will open a new window at very low frequencies in the KHz--MHz range \cite{2017AdSpR..59..736B}. These
satellites are expected to function as a distributed low frequency array far away from the blocking ionosphere and
terrestrial radio frequency interference, which will have stable conditions for observing the cosmic signals.
Such low frequency observations would offer a more sensitive probe of any delays expected from a nonzero photon mass.

(iii) Shapiro delay-based tests of the WEP have reached high precision.

Pioneered by SN 1987A tests \cite{1988PhRvL..60..173L,1988PhRvL..60..176K}, the Shapiro delay experienced by an astrophysical
messenger traveling through a gravitational field has been intensively employed to constrain possible violations of the WEP
(e.g., \cite{2015ApJ...810..121G,2015PhRvL.115z1101W,2016PhRvD..94b4061W,2017MNRAS.469L..36Y,2019PhRvD..99j3012W}).
However, there are some uncertainties and caveats involved in such tests. The large uncertainty is from the estimation
of the local gravitational potential $U(r)$. The exact gravitational potential function is not well known. More accurate models of the function $U(r)$ could improve the constraints on the WEP violation, but the improvement should be very limited.
Most often the Shapiro delay terms caused by the host galaxy and the intergalactic gravitational potential are ignored.
In principle, these terms might be much larger than that caused by the local gravitational potential. With the better
understanding of the potential functions for these terms, the WEP tests might be improved by orders of magnitude.
As explained in Gao \&  Wald \cite{2000CQGra..17.4999G}, Eq.~(\ref{eq:Shapiro}) is gauge dependent, i.e.,
depending on the coordinate choice one can obtain both positive and negative values of the Shapiro delay.
Additionally, Eq.~(\ref{eq:Shapiro}) implicitly assumes that the trajectory is short enough that one can treat the cosmological
spacetime as Minkowski plus a linear perturbation. Minazzoli et al. \cite{2019PhRvD.100j4047M} showed that this assumption
is well justified for sufficiently nearby sources like GW170817, but not for sources at cosmological distances such
as GRBs or FRBs with redshifts $z\geq1$. While Nusser \cite{2016ApJ...821L...2N} has provided a formulation that, in principle,
can be applied to more distant sources, most of the previously cited works use the standard expression for the Shapiro delay
(Eq.~\ref{eq:Shapiro}). This is an additional source of uncertainty involved in such WEP tests.
The standard use of Eq.~(\ref{eq:Shapiro}) also suffers from an assumption that there exists a coordinate time such that
the potential and its derivative vanishes at infinity. Minazzoli et al. \cite{2019PhRvD.100j4047M} showed that such
an assumption results in a spurious divergence of the Shapiro delay for increasingly remote sources. With the use
of an adequate coordinate time, Minazzoli et al. \cite{2019PhRvD.100j4047M} found that the Shapiro delay is no longer
monotonic with the number of the sources of the gravitational field. Hence, without further assumptions and/or observational input, one can not obtain a conservative lower limit on the Shapiro delay from a subset of the gravitational sources based on
Eq.~(\ref{eq:Shapiro}). It might be possible to constrain the gravitational potential along the line of sight using
cosmological observations (e.g., of galaxy peculiar velocities, as in Ref. \cite{2017NatAs...1E..36H}), and thus avoid
the need to use Eq.~(\ref{eq:Shapiro}). Minazzoli et al. \cite{2019PhRvD.100j4047M} suggested that any further developments
of this Shapiro delay-based test should be inspired by a fundamental theory to avoid the gauge dependence of the current
expression of the test.

\acknowledgements{We are grateful to the anonymous referees for insightful comments.
This work is partially supported by the National Natural Science Foundation of China
(grant Nos. 11673068, 11725314, U1831122, and 12041306), the Youth Innovation Promotion
Association (2017366), the Key Research Program of Frontier Sciences (grant Nos. QYZDB-SSW-SYS005
and ZDBS-LY-7014), and the Strategic Priority Research Program ``Multi-waveband gravitational wave universe''
(grant No. XDB23000000) of Chinese Academy of Sciences.
}


\begin{thebibliography}{100}
\providecommand{\url}[1]{{#1}}
\providecommand{\urlprefix}{URL }
\expandafter\ifx\csname urlstyle\endcsname\relax
  \providecommand{\doi}[1]{DOI \discretionary{}{}{}#1}\else
  \providecommand{\doi}{DOI \discretionary{}{}{}\begingroup
  \urlstyle{rm}\Url}\fi

\bibitem{1999PhRvD..59l4021G}
R.~{Gambini}, J.~{Pullin}, \prd \textbf{59}(12), 124021 (1999).
\newblock \doi{10.1103/PhysRevD.59.124021}

\bibitem{2002PhRvD..65j3509A}
J.~{Alfaro}, H.A. {Morales-T{\'e}cotl}, L.F. {Urrutia}, \prd \textbf{65}(10),
  103509 (2002).
\newblock \doi{10.1103/PhysRevD.65.103509}

\bibitem{2002IJMPD..11...35A}
G.~{Amelino-Camelia}, D.V. {Ahluwalia}, International Journal of Modern Physics
  D \textbf{11}(1), 35 (2002).
\newblock \doi{10.1142/S0218271802001330}

\bibitem{2002Natur.418...34A}
G.~{Amelino-Camelia}, \nat \textbf{418}(6893), 34 (2002).
\newblock \doi{10.1038/418034a}

\bibitem{2002PhLB..539..126K}
J.~{Kowalski-Glikman}, S.~{Nowak}, Physics Letters B \textbf{539}(1), 126
  (2002).
\newblock \doi{10.1016/S0370-2693(02)02063-4}

\bibitem{2003PhRvD..67d4017M}
J.~{Magueijo}, L.~{Smolin}, \prd \textbf{67}(4), 044017 (2003).
\newblock \doi{10.1103/PhysRevD.67.044017}

\bibitem{1989PhRvD..39..683K}
V.A. {Kosteleck{\'y}}, S.~{Samuel}, \prd \textbf{39}, 683 (1989).
\newblock \doi{10.1103/PhysRevD.39.683}

\bibitem{1991NuPhB.359..545K}
V.A. {Kosteleck{\'y}}, R.~{Potting}, Nuclear Physics B \textbf{359}, 545 (1991)

\bibitem{1995PhRvD..51.3923K}
V.A. {Kosteleck{\'y}}, R.~{Potting}, \prd \textbf{51}, 3923 (1995).
\newblock \doi{10.1103/PhysRevD.51.3923}

\bibitem{2005LRR.....8....5M}
D.~{Mattingly}, Living Reviews in Relativity \textbf{8}, 5 (2005).
\newblock \doi{10.12942/lrr-2005-5}

\bibitem{2005hep.ph....6054B}
R.~{Bluhm}, Lecture Notes in Physics \textbf{702}, 191 (2006).
\newblock \doi{10.1007/3-540-34523-X\_8}

\bibitem{2013LRR....16....5A}
G.~{Amelino-Camelia}, Living Reviews in Relativity \textbf{16}, 5 (2013).
\newblock \doi{10.12942/lrr-2013-5}

\bibitem{2014RPPh...77f2901T}
J.D. {Tasson}, Reports on Progress in Physics \textbf{77}(6), 062901 (2014).
\newblock \doi{10.1088/0034-4885/77/6/062901}

\bibitem{2011RvMP...83...11K}
V.A. {Kosteleck{\'y}}, N.~{Russell}, Reviews of Modern Physics \textbf{83}, 11
  (2011).
\newblock \doi{10.1103/RevModPhys.83.11}

\bibitem{2008ApJ...689L...1K}
V.A. {Kosteleck{\'y}}, M.~{Mewes}, \apjl \textbf{689}, L1 (2008).
\newblock \doi{10.1086/595815}

\bibitem{1998Natur.393..763A}
G.~{Amelino-Camelia}, J.~{Ellis}, N.E. {Mavromatos}, D.V. {Nanopoulos},
  S.~{Sarkar}, \nat \textbf{393}, 763 (1998).
\newblock \doi{10.1038/31647}

\bibitem{2005PhLB..625...13P}
T.G. {Pavlopoulos}, Physics Letters B \textbf{625}, 13 (2005).
\newblock \doi{10.1016/j.physletb.2005.08.064}

\bibitem{2006APh....25..402E}
J.~{Ellis}, N.E. {Mavromatos}, D.V. {Nanopoulos}, A.S. {Sakharov}, E.K.G.
  {Sarkisyan}, Astroparticle Physics \textbf{25}, 402 (2006).
\newblock \doi{10.1016/j.astropartphys.2006.04.001}

\bibitem{2008JCAP...01..031J}
U.~{Jacob}, T.~{Piran}, \jcap \textbf{1}, 031 (2008).
\newblock \doi{10.1088/1475-7516/2008/01/031}

\bibitem{2009PhRvD..80a5020K}
V.A. {Kosteleck{\'y}}, M.~{Mewes}, \prd \textbf{80}(1), 015020 (2009).
\newblock \doi{10.1103/PhysRevD.80.015020}

\bibitem{2009Sci...323.1688A}
A.A. {Abdo}, M.~{Ackermann}, M.~{Arimoto}, et al., Science \textbf{323}(5922), 1688 (2009).
\newblock \doi{10.1126/science.1169101}

\bibitem{2009Natur.462..331A}
A.A. {Abdo}, M.~{Ackermann}, M.~{Ajello}, et al., \nat
  \textbf{462}(7271), 331 (2009).
\newblock \doi{10.1038/nature08574}

\bibitem{2012APh....36...47C}
Z.~{Chang}, Y.~{Jiang}, H.N. {Lin}, Astroparticle Physics \textbf{36}(1), 47
  (2012).
\newblock \doi{10.1016/j.astropartphys.2012.04.006}

\bibitem{2012PhRvL.108w1103N}
R.J. {Nemiroff}, R.~{Connolly}, J.~{Holmes}, A.B. {Kostinski}, \prl
  \textbf{108}(23), 231103 (2012).
\newblock \doi{10.1103/PhysRevLett.108.231103}

\bibitem{2013PhRvD..87l2001V}
V.~{Vasileiou}, A.~{Jacholkowska}, F.~{Piron}, J.~{Bolmont}, C.~{Couturier},
  J.~{Granot}, F.W. {Stecker}, J.~{Cohen-Tanugi}, F.~{Longo}, \prd
  \textbf{87}(12), 122001 (2013).
\newblock \doi{10.1103/PhysRevD.87.122001}

\bibitem{2013APh....43...50E}
J.~{Ellis}, N.E. {Mavromatos}, Astroparticle Physics \textbf{43}, 50 (2013).
\newblock \doi{10.1016/j.astropartphys.2012.05.004}

\bibitem{2015PhRvD..92d5016K}
F.~{Kislat}, H.~{Krawczynski}, \prd \textbf{92}(4), 045016 (2015).
\newblock \doi{10.1103/PhysRevD.92.045016}

\bibitem{2015APh....61..108Z}
S.~{Zhang}, B.Q. {Ma}, Astroparticle Physics \textbf{61}, 108 (2015).
\newblock \doi{10.1016/j.astropartphys.2014.04.008}

\bibitem{2017ApJ...834L..13W}
J.J. {Wei}, B.B. {Zhang}, L.~{Shao}, X.F. {Wu}, P.~{M{\'e}sz{\'a}ros}, \apjl
  \textbf{834}(2), L13 (2017).
\newblock \doi{10.3847/2041-8213/834/2/L13}

\bibitem{2017ApJ...842..115W}
J.J. {Wei}, X.F. {Wu}, B.B. {Zhang}, L.~{Shao}, P.~{M{\'e}sz{\'a}ros}, V.A.
  {Kosteleck{\'y}}, \apj \textbf{842}(2), 115 (2017).
\newblock \doi{10.3847/1538-4357/aa7630}

\bibitem{2017ApJ...851..127W}
J.J. {Wei}, X.F. {Wu}, \apj \textbf{851}(2), 127 (2017).
\newblock \doi{10.3847/1538-4357/aa9d8d}

\bibitem{2019PhRvD..99h3009E}
J.~{Ellis}, R.~{Konoplich}, N.E. {Mavromatos}, L.~{Nguyen}, A.S. {Sakharov},
  E.K. {Sarkisyan-Grinbaum}, \prd \textbf{99}(8), 083009 (2019).
\newblock \doi{10.1103/PhysRevD.99.083009}

\bibitem{PhysRevLett.125.021301}
V.A. Acciari, S.~Ansoldi, L.A. Antonelli, et al., Phys. Rev. Lett.
  \textbf{125}, 021301 (2020).
\newblock \doi{10.1103/PhysRevLett.125.021301}.

\bibitem{1999PhRvL..83.2108B}
S.D. {Biller}, A.C. {Breslin}, J.~{Buckley}, et al., \prl
  \textbf{83}(11), 2108 (1999).
\newblock \doi{10.1103/PhysRevLett.83.2108}

\bibitem{Kaaret1999}
P.~{Kaaret}, \aap \textbf{345}, L32 (1999)

\bibitem{1990PhRvD..41.1231C}
S.M. {Carroll}, G.B. {Field}, R.~{Jackiw}, \prd \textbf{41}(4), 1231 (1990).
\newblock \doi{10.1103/PhysRevD.41.1231}

\bibitem{1998PhRvD..58k6002C}
D.~{Colladay}, V.A. {Kosteleck{\'y}}, \prd \textbf{58}(11), 116002 (1998).
\newblock \doi{10.1103/PhysRevD.58.116002}

\bibitem{2001PhRvD..64h3007G}
R.J. {Gleiser}, C.N. {Kozameh}, \prd \textbf{64}(8), 083007 (2001).
\newblock \doi{10.1103/PhysRevD.64.083007}

\bibitem{2001PhRvL..87y1304K}
V.A. {Kosteleck{\'y}}, M.~{Mewes}, Physical Review Letters \textbf{87}(25),
  251304 (2001).
\newblock \doi{10.1103/PhysRevLett.87.251304}

\bibitem{2006PhRvL..97n0401K}
V.A. {Kosteleck{\'y}}, M.~{Mewes}, Physical Review Letters \textbf{97}(14),
  140401 (2006).
\newblock \doi{10.1103/PhysRevLett.97.140401}

\bibitem{2007PhRvL..99a1601K}
V.A. {Kosteleck{\'y}}, M.~{Mewes}, Physical Review Letters \textbf{99}(1),
  011601 (2007).
\newblock \doi{10.1103/PhysRevLett.99.011601}

\bibitem{2013PhRvL.110t1601K}
V.A. {Kosteleck{\'y}}, M.~{Mewes}, Physical Review Letters \textbf{110}(20),
  201601 (2013).
\newblock \doi{10.1103/PhysRevLett.110.201601}

\bibitem{2003Natur.426Q.139M}
I.G. {Mitrofanov}, \nat \textbf{426}, 139 (2003).
\newblock \doi{10.1038/426139a}

\bibitem{2004PhRvL..93b1101J}
T.~{Jacobson}, S.~{Liberati}, D.~{Mattingly}, F.W. {Stecker}, Physical Review
  Letters \textbf{93}(2), 021101 (2004).
\newblock \doi{10.1103/PhysRevLett.93.021101}

\bibitem{2007MNRAS.376.1857F}
Y.Z. {Fan}, D.M. {Wei}, D.~{Xu}, \mnras \textbf{376}, 1857 (2007).
\newblock \doi{10.1111/j.1365-2966.2007.11576.x}

\bibitem{2009JCAP...08..021G}
G.~{Gubitosi}, L.~{Pagano}, G.~{Amelino-Camelia}, A.~{Melchiorri}, A.~{Cooray},
  \jcap \textbf{8}, 021 (2009).
\newblock \doi{10.1088/1475-7516/2009/08/021}

\bibitem{2011PhRvD..83l1301L}
P.~{Laurent}, D.~{G{\"o}tz}, P.~{Bin{\'e}truy}, S.~{Covino},
  A.~{Fernandez-Soto}, \prd \textbf{83}(12), 121301 (2011).
\newblock \doi{10.1103/PhysRevD.83.121301}

\bibitem{2011APh....35...95S}
F.W. {Stecker}, Astroparticle Physics \textbf{35}, 95 (2011).
\newblock \doi{10.1016/j.astropartphys.2011.06.007}

\bibitem{2012PhRvL.109x1104T}
K.~{Toma}, S.~{Mukohyama}, D.~{Yonetoku}, T.~{Murakami}, S.~{Gunji},
  T.~{Mihara}, Y.~{Morihara}, T.~{Sakashita}, T.~{Takahashi}, Y.~{Wakashima},
  H.~{Yonemochi}, N.~{Toukairin}, Physical Review Letters \textbf{109}(24),
  241104 (2012).
\newblock \doi{10.1103/PhysRevLett.109.241104}

\bibitem{2013MNRAS.431.3550G}
D.~{G{\"o}tz}, S.~{Covino}, A.~{Fern{\'a}ndez-Soto}, P.~{Laurent}, {\v
  Z}.~{Bo{\v s}njak}, \mnras \textbf{431}, 3550 (2013).
\newblock \doi{10.1093/mnras/stt439}

\bibitem{2014MNRAS.444.2776G}
D.~{G{\"o}tz}, P.~{Laurent}, S.~{Antier}, S.~{Covino}, P.~{D'Avanzo},
  V.~{D'Elia}, A.~{Melandri}, \mnras \textbf{444}, 2776 (2014).
\newblock \doi{10.1093/mnras/stu1634}

\bibitem{2016MNRAS.463..375L}
H.N. {Lin}, X.~{Li}, Z.~{Chang}, \mnras \textbf{463}, 375 (2016).
\newblock \doi{10.1093/mnras/stw2007}

\bibitem{2017PhRvD..95h3013K}
F.~{Kislat}, H.~{Krawczynski}, \prd \textbf{95}(8), 083013 (2017).
\newblock \doi{10.1103/PhysRevD.95.083013}

\bibitem{2019PhRvD..99c5045F}
A.S. {Friedman}, D.~{Leon}, K.D. {Crowley}, D.~{Johnson}, G.~{Teply},
  D.~{Tytler}, B.G. {Keating}, G.M. {Cole}, \prd \textbf{99}(3), 035045 (2019).
\newblock \doi{10.1103/PhysRevD.99.035045}

\bibitem{2019MNRAS.485.2401W}
J.J. {Wei}, \mnras \textbf{485}(2), 2401 (2019).
\newblock \doi{10.1093/mnras/stz594}

\bibitem{1971RvMP...43..277G}
A.S. {Goldhaber}, M.M. {Nieto}, Reviews of Modern Physics \textbf{43}(3), 277
  (1971).
\newblock \doi{10.1103/RevModPhys.43.277}

\bibitem{2005RPPh...68...77T}
L.C. {Tu}, J.~{Luo}, G.T. {Gillies}, Reports on Progress in Physics
  \textbf{68}(1), 77 (2005).
\newblock \doi{10.1088/0034-4885/68/1/R02}

\bibitem{2006AcPPB..37..565O}
L.B. {Okun}, Acta Physica Polonica B \textbf{37}(3), 565 (2006)

\bibitem{2010RvMP...82..939G}
A.S. {Goldhaber}, M.M. {Nieto}, Reviews of Modern Physics \textbf{82}(1), 939
  (2010).
\newblock \doi{10.1103/RevModPhys.82.939}

\bibitem{2011EPJD...61..531S}
G.~{Spavieri}, J.~{Quintero}, G.T. {Gillies}, M.~{Rodr{\'\i}guez}, European
  Physical Journal D \textbf{61}(3), 531 (2011).
\newblock \doi{10.1140/epjd/e2011-10508-7}

\bibitem{1964Natur.202..377L}
B.~{Lovell}, F.L. {Whipple}, L.H. {Solomon}, \nat \textbf{202}(4930), 377
  (1964).
\newblock \doi{10.1038/202377a0}

\bibitem{1969Natur.222..157W}
B.~{Warner}, R.E. {Nather}, \nat \textbf{222}(5189), 157 (1969).
\newblock \doi{10.1038/222157b0}

\bibitem{1999PhRvL..82.4964S}
B.E. {Schaefer}, \prl \textbf{82}(25), 4964 (1999).
\newblock \doi{10.1103/PhysRevLett.82.4964}

\bibitem{2016JHEAp..11...20Z}
B.~{Zhang}, Y.T. {Chai}, Y.C. {Zou}, X.F. {Wu}, Journal of High Energy
  Astrophysics \textbf{11}, 20 (2016).
\newblock \doi{10.1016/j.jheap.2016.07.001}

\bibitem{2017RAA....17...13W}
J.J. {Wei}, E.K. {Zhang}, S.B. {Zhang}, X.F. {Wu}, Research in Astronomy and
  Astrophysics \textbf{17}(2), 13 (2017).
\newblock \doi{10.1088/1674-4527/17/2/13}

\bibitem{2016ApJ...822L..15W}
X.F. {Wu}, S.B. {Zhang}, H.~{Gao}, J.J. {Wei}, Y.C. {Zou}, W.H. {Lei},
  B.~{Zhang}, Z.G. {Dai}, P.~{M{\'e}sz{\'a}ros}, \apjl \textbf{822}(1), L15
  (2016).
\newblock \doi{10.3847/2041-8205/822/1/L15}

\bibitem{2016PhLB..757..548B}
L.~{Bonetti}, J.~{Ellis}, N.E. {Mavromatos}, A.S. {Sakharov}, E.K.
  {Sarkisyan-Grinbaum}, A.D.A.M. {Spallicci}, Physics Letters B \textbf{757},
  548 (2016).
\newblock \doi{10.1016/j.physletb.2016.04.035}

\bibitem{2017PhLB..768..326B}
L.~{Bonetti}, J.~{Ellis}, N.E. {Mavromatos}, A.S. {Sakharov}, E.K.
  {Sarkisyan-Grinbaum}, A.D.A.M. {Spallicci}, Physics Letters B \textbf{768},
  326 (2017).
\newblock \doi{10.1016/j.physletb.2017.03.014}

\bibitem{2017PhRvD..95l3010S}
L.~{Shao}, B.~{Zhang}, \prd \textbf{95}(12), 123010 (2017).
\newblock \doi{10.1103/PhysRevD.95.123010}

\bibitem{2018JCAP...07..045W}
J.J. {Wei}, X.F. {Wu}, \jcap \textbf{2018}(7), 045 (2018).
\newblock \doi{10.1088/1475-7516/2018/07/045}

\bibitem{2019ApJ...882L..13X}
N.~{Xing}, H.~{Gao}, J.J. {Wei}, Z.~{Li}, W.~{Wang}, B.~{Zhang}, X.F. {Wu},
  P.~{M{\'e}sz{\'a}ros}, \apjl \textbf{882}(1), L13 (2019).
\newblock \doi{10.3847/2041-8213/ab3c5f}

\bibitem{2020arXiv200609680W}
J.J. {Wei}, X.F. {Wu}, Research in Astronomy and Astrophysics \textbf{20}(12), 206 (2020).
\newblock \doi{10.1088/1674-4527/20/12/206}

\bibitem{1971PhRvL..26..721W}
E.R. {Williams}, J.E. {Faller}, H.A. {Hill}, \prl \textbf{26}(12), 721 (1971).
\newblock \doi{10.1103/PhysRevLett.26.721}

\bibitem{1992PhRvL..68.3383C}
M.A. {Chernikov}, C.J. {Gerber}, H.R. {Ott}, H.J. {Gerber}, \prl
  \textbf{68}(23), 3383 (1992).
\newblock \doi{10.1103/PhysRevLett.68.3383}

\bibitem{1998PhRvL..80.1826L}
R.~{Lakes}, \prl \textbf{80}(9), 1826 (1998).
\newblock \doi{10.1103/PhysRevLett.80.1826}

\bibitem{2003PhRvL..91n9101G}
A.S. {Goldhaber}, M.M. {Nieto}, \prl \textbf{91}(14), 149101 (2003).
\newblock \doi{10.1103/PhysRevLett.91.149101}

\bibitem{2003PhRvL..90h1801L}
J.~{Luo}, L.C. {Tu}, Z.K. {Hu}, E.J. {Luan}, \prl \textbf{90}(8), 081801
  (2003).
\newblock \doi{10.1103/PhysRevLett.90.081801}

\bibitem{2003PhRvL..91n9102L}
J.~{Luo}, L.C. {Tu}, Z.K. {Hu}, E.J. {Luan}, \prl \textbf{91}(14), 149102
  (2003).
\newblock \doi{10.1103/PhysRevLett.91.149102}

\bibitem{1973PhRvD...8.2349L}
D.D. {Lowenthal}, \prd \textbf{8}(8), 2349 (1973).
\newblock \doi{10.1103/PhysRevD.8.2349}

\bibitem{2004PhRvD..69j7501A}
A.~{Accioly}, R.~{Paszko}, \prd \textbf{69}(10), 107501 (2004).
\newblock \doi{10.1103/PhysRevD.69.107501}

\bibitem{1975PhRvL..35.1402D}
J.~{Davis}, L., A.S. {Goldhaber}, M.M. {Nieto}, \prl \textbf{35}(21), 1402
  (1975).
\newblock \doi{10.1103/PhysRevLett.35.1402}

\bibitem{1997PPCF...39...73R}
D.D. {Ryutov}, Plasma Physics and Controlled Fusion \textbf{39}, A73 (1997).
\newblock \doi{10.1088/0741-3335/39/5A/008}

\bibitem{2007PPCF...49..429R}
D.D. {Ryutov}, Plasma Physics and Controlled Fusion \textbf{49}(12B), B429
  (2007).
\newblock \doi{10.1088/0741-3335/49/12B/S40}

\bibitem{2016APh....82...49R}
A.~{Retin{\`o}}, A.D.A.M. {Spallicci}, A.~{Vaivads}, Astroparticle Physics
  \textbf{82}, 49 (2016).
\newblock \doi{10.1016/j.astropartphys.2016.05.006}

\bibitem{1959PThPS..11....1Y}
Y.~{Yamaguchi}, Progress of Theoretical Physics Supplement \textbf{11}, 1
  (1959).
\newblock \doi{10.1143/PTPS.11.1}

\bibitem{1976UsFiN.119..551C}
G.V. {Chibisov}, Uspekhi Fizicheskikh Nauk \textbf{119}, 551 (1976)

\bibitem{PhysRevLett.98.010402}
E.~Adelberger, G.~Dvali, A.~Gruzinov, Phys. Rev. Lett. \textbf{98}, 010402
  (2007).
\newblock \doi{10.1103/PhysRevLett.98.010402}.

\bibitem{2012PhRvL.109m1102P}
P.~{Pani}, V.~{Cardoso}, L.~{Gualtieri}, E.~{Berti}, A.~{Ishibashi}, \prl
  \textbf{109}(13), 131102 (2012).
\newblock \doi{10.1103/PhysRevLett.109.131102}

\bibitem{2017ApJ...842...23Y}
Y.P. {Yang}, B.~{Zhang}, \apj \textbf{842}(1), 23 (2017).
\newblock \doi{10.3847/1538-4357/aa74de}

\bibitem{2006LRR.....9....3W}
C.M. {Will}, Living Rev. Rel. \textbf{9}, 3 (2006).
\newblock \doi{10.12942/lrr-2006-3}

\bibitem{2014LRR....17....4W}
C.M. {Will}, Living Rev. Rel. \textbf{17}, 4 (2014).
\newblock \doi{10.12942/lrr-2014-4}

\bibitem{2009A&A...499..331L}
S.B. {Lambert}, C.~{Le Poncin-Lafitte}, \aap \textbf{499}(1), 331 (2009).
\newblock \doi{10.1051/0004-6361/200911714}

\bibitem{2011A&A...529A..70L}
S.B. {Lambert}, C.~{Le Poncin-Lafitte}, \aap \textbf{529}, A70 (2011).
\newblock \doi{10.1051/0004-6361/201016370}

\bibitem{2003Natur.425..374B}
B.~{Bertotti}, L.~{Iess}, P.~{Tortora}, \nat \textbf{425}(6956), 374 (2003).
\newblock \doi{10.1038/nature01997}

\bibitem{1964PhRvL..13..789S}
I.I. {Shapiro}, Phys. Rev. Lett. \textbf{13}, 789 (1964).
\newblock \doi{10.1103/PhysRevLett.13.789}

\bibitem{1988PhRvL..60..173L}
M.J. {Longo}, Phys. Rev. Lett. \textbf{60}, 173 (1988).
\newblock \doi{10.1103/PhysRevLett.60.173}

\bibitem{1988PhRvL..60..176K}
L.M. {Krauss}, S.~{Tremaine}, Phys. Rev. Lett. \textbf{60}, 176 (1988).
\newblock \doi{10.1103/PhysRevLett.60.176}

\bibitem{2015ApJ...810..121G}
H.~{Gao}, X.F. {Wu}, P.~{M{\'e}sz{\'a}ros}, Astrophys. J. \textbf{810}, 121
  (2015).
\newblock \doi{10.1088/0004-637X/810/2/121}

\bibitem{2015PhRvL.115z1101W}
J.J. {Wei}, H.~{Gao}, X.F. {Wu}, P.~{M{\'e}sz{\'a}ros}, Phys. Rev. Lett.
  \textbf{115}(26), 261101 (2015).
\newblock \doi{10.1103/PhysRevLett.115.261101}

\bibitem{2016PhRvD..94b4061W}
X.F. {Wu}, H.~{Gao}, J.J. {Wei}, P.~{M{\'e}sz{\'a}ros}, B.~{Zhang}, Z.G. {Dai},
  S.N. {Zhang}, Z.H. {Zhu}, Phys. Rev. D \textbf{94}(2), 024061 (2016).
\newblock \doi{10.1103/PhysRevD.94.024061}

\bibitem{2017MNRAS.469L..36Y}
C.~{Yang}, Y.C. {Zou}, Y.Y. {Zhang}, B.~{Liao}, W.H. {Lei}, Mon. Not. R.
  Astron. Soc. \textbf{469}, L36 (2017).
\newblock \doi{10.1093/mnrasl/slx045}

\bibitem{2019PhRvD..99j3012W}
J.J. {Wei}, X.F. {Wu}, \prd \textbf{99}(10), 103012 (2019).
\newblock \doi{10.1103/PhysRevD.99.103012}

\bibitem{2003hep.th....3185S}
L.~{Smolin}, arXiv e-prints hep-th/0303185 (2003)

\bibitem{1998LRR.....1....1R}
C.~{Rovelli}, Living Reviews in Relativity \textbf{1}(1), 1 (1998).
\newblock \doi{10.12942/lrr-1998-1}

\bibitem{2019arXiv191102154B}
L.~{Burderi}, A.~{Sanna}, T.~{Di Salvo}, L.~{Amati}, G.~{Amelino-Camelia},
  M.~{Branchesi}, S.~{Capozziello}, E.~{Coccia}, M.~{Colpi}, E.~{Costa},
  N.~{D'Amico}, P.~{De Bernardis}, M.~{De Laurentis}, M.~{Della Valle},
  H.~{Falcke}, M.~{Feroci}, F.~{Fiore}, F.~{Frontera}, A.F. {Gambino},
  G.~{Ghisellini}, K.~{Hurley}, R.~{Iaria}, D.~{Kataria}, C.~{Labanti},
  G.~{Lodato}, B.~{Negri}, A.~{Papitto}, T.~{Piran}, A.~{Riggio}, C.~{Rovelli},
  A.~{Santangelo}, F.~{Vidotto}, S.~{Zane}, arXiv e-prints arXiv:1911.02154
  (2019)

\bibitem{2008APh....29..158E}
J.~{Ellis}, N.E. {Mavromatos}, D.V. {Nanopoulos}, A.S. {Sakharov}, E.K.G.
  {Sarkisyan}, Astroparticle Physics \textbf{29}(2), 158 (2008).
\newblock \doi{10.1016/j.astropartphys.2007.12.003}

\bibitem{Ellis2003}
J.~{Ellis}, N.E. {Mavromatos}, D.V. {Nanopoulos}, A.S. {Sakharov}, \aap
  \textbf{402}, 409 (2003).
\newblock \doi{10.1051/0004-6361:20030263}

\bibitem{2004ApJ...611L..77B}
S.E. {Boggs}, C.B. {Wunderer}, K.~{Hurley}, W.~{Coburn}, \apjl \textbf{611}(2),
  L77 (2004).
\newblock \doi{10.1086/423933}

\bibitem{2006JCAP...05..017R}
M.~{Rodr{\'\i}guez Mart{\'\i}nez}, T.~{Piran}, Y.~{Oren}, \jcap
  \textbf{2006}(5), 017 (2006).
\newblock \doi{10.1088/1475-7516/2006/05/017}

\bibitem{2008ApJ...676..532B}
J.~{Bolmont}, A.~{Jacholkowska}, J.L. {Atteia}, F.~{Piron}, G.~{Pizzichini},
  \apj \textbf{676}(1), 532 (2008).
\newblock \doi{10.1086/527524}

\bibitem{2008GReGr..40.1731L}
R.~{Lamon}, N.~{Produit}, F.~{Steiner}, General Relativity and Gravitation
  \textbf{40}(8), 1731 (2008).
\newblock \doi{10.1007/s10714-007-0580-6}

\bibitem{2009PhRvD..80k6005X}
Z.~{Xiao}, B.Q. {Ma}, \prd \textbf{80}(11), 116005 (2009).
\newblock \doi{10.1103/PhysRevD.80.116005}

\bibitem{2010APh....33..312S}
L.~{Shao}, Z.~{Xiao}, B.Q. {Ma}, Astroparticle Physics \textbf{33}(5-6), 312
  (2010).
\newblock \doi{10.1016/j.astropartphys.2010.03.003}

\bibitem{2016APh....82...72X}
H.~{Xu}, B.Q. {Ma}, Astroparticle Physics \textbf{82}, 72 (2016).
\newblock \doi{10.1016/j.astropartphys.2016.05.008}

\bibitem{2016PhLB..760..602X}
H.~{Xu}, B.Q. {Ma}, Physics Letters B \textbf{760}, 602 (2016).
\newblock \doi{10.1016/j.physletb.2016.07.044}

\bibitem{2018JCAP...01..050X}
H.~{Xu}, B.Q. {Ma}, \jcap \textbf{2018}(1), 050 (2018).
\newblock \doi{10.1088/1475-7516/2018/01/050}

\bibitem{Liu2018}
Y.~Liu, B.Q. Ma, The European Physical Journal C \textbf{78}(10), 825 (2018).
\newblock \doi{10.1140/epjc/s10052-018-6294-y}.

\bibitem{2019Natur.575..455M}
{MAGIC Collaboration}, V.A. {Acciari}, S.~{Ansoldi}, et al., \nat \textbf{575}(7783), 455
  (2019).
\newblock \doi{10.1038/s41586-019-1750-x}

\bibitem{2009CQGra..26l5007B}
M.~{Biesiada}, A.~{Pi{\'o}rkowska}, Classical and Quantum Gravity
  \textbf{26}(12), 125007 (2009).
\newblock \doi{10.1088/0264-9381/26/12/125007}

\bibitem{2015ApJ...808...78P}
Y.~{Pan}, Y.~{Gong}, S.~{Cao}, H.~{Gao}, Z.H. {Zhu}, \apj \textbf{808}(1), 78
  (2015).
\newblock \doi{10.1088/0004-637X/808/1/78}

\bibitem{2018PhLB..776..284Z}
X.B. {Zou}, H.K. {Deng}, Z.Y. {Yin}, H.~{Wei}, Physics Letters B \textbf{776},
  284 (2018).
\newblock \doi{10.1016/j.physletb.2017.11.053}

\bibitem{2020ApJ...890..169P}
Y.~{Pan}, J.~{Qi}, S.~{Cao}, T.~{Liu}, Y.~{Liu}, S.~{Geng}, Y.~{Lian}, Z.H.
  {Zhu}, \apj \textbf{890}(2), 169 (2020).
\newblock \doi{10.3847/1538-4357/ab6ef5}

\bibitem{2012MNRAS.419..614U}
T.N. {Ukwatta}, K.S. {Dhuga}, M.~{Stamatikos}, C.D. {Dermer}, T.~{Sakamoto},
  E.~{Sonbas}, W.C. {Parke}, L.C. {Maximon}, J.T. {Linnemann}, P.N. {Bhat},
  A.~{Eskand arian}, N.~{Gehrels}, A.U. {Abeysekara}, K.~{Tollefson}, J.P.
  {Norris}, \mnras \textbf{419}(1), 614 (2012).
\newblock \doi{10.1111/j.1365-2966.2011.19723.x}

\bibitem{2015MNRAS.446.1129B}
M.G. {Bernardini}, G.~{Ghirlanda}, S.~{Campana}, S.~{Covino}, R.~{Salvaterra},
  J.L. {Atteia}, D.~{Burlon}, G.~{Calderone}, P.~{D'Avanzo}, V.~{D'Elia},
  G.~{Ghisellini}, V.~{Heussaff}, D.~{Lazzati}, A.~{Meland ri}, L.~{Nava}, S.D.
  {Vergani}, G.~{Tagliaferri}, \mnras \textbf{446}(2), 1129 (2015).
\newblock \doi{10.1093/mnras/stu2153}

\bibitem{2016ChPhC..40d5102C}
Z.~{Chang}, X.~{Li}, H.N. {Lin}, Y.~{Sang}, P.~{Wang}, S.~{Wang}, Chinese
  Physics C \textbf{40}(4), 045102 (2016).
\newblock \doi{10.1088/1674-1137/40/4/045102}

\bibitem{2017ApJ...844..126S}
L.~{Shao}, B.B. {Zhang}, F.R. {Wang}, X.F. {Wu}, Y.H. {Cheng}, X.~{Zhang}, B.Y.
  {Yu}, B.J. {Xi}, X.~{Wang}, H.X. {Feng}, M.~{Zhang}, D.~{Xu}, \apj
  \textbf{844}(2), 126 (2017).
\newblock \doi{10.3847/1538-4357/aa7d01}

\bibitem{2018ApJ...865..153L}
R.J. {Lu}, Y.F. {Liang}, D.B. {Lin}, J.~{L{\"u}}, X.G. {Wang}, H.J. {L{\"u}},
  H.B. {Liu}, E.W. {Liang}, B.~{Zhang}, \apj \textbf{865}(2), 153 (2018).
\newblock \doi{10.3847/1538-4357/aada16}

\bibitem{2008PhLB..668..253M}
{MAGIC Collaboration}, J.~{Albert}, E.~{Aliu}, et al., Physics Letters B \textbf{668}, 253 (2008).
\newblock \doi{10.1016/j.physletb.2008.08.053}

\bibitem{2009APh....31..226M}
M.~{Mart{\'\i}nez}, M.~{Errando}, Astroparticle Physics \textbf{31}(3), 226
  (2009).
\newblock \doi{10.1016/j.astropartphys.2009.01.005}

\bibitem{2019ApJ...870...93A}
H.~{Abdalla}, F.~{Aharonian}, F.~{Ait Benkhali}, et al., \apj \textbf{870}(2), 93 (2019).
\newblock \doi{10.3847/1538-4357/aaf1c4}

\bibitem{2008PhRvL.101q0402A}
F.~{Aharonian}, A.G. {Akhperjanian}, U.~{Barres de Almeida}, et al., Physical Review Letters
  \textbf{101}(17), 170402 (2008).
\newblock \doi{10.1103/PhysRevLett.101.170402}

\bibitem{2011APh....34..738H}
{H.~E.~S.~S. Collaboration}, A.~{Abramowski}, F.~{Acero}, et al.,
  Astroparticle Physics \textbf{34}(9), 738 (2011).
\newblock \doi{10.1016/j.astropartphys.2011.01.007}

\bibitem{2011ICRC....7..256O}
N.~{OTTE}, in \emph{International Cosmic Ray Conference}, \emph{International
  Cosmic Ray Conference}, vol.~7 (2011), \emph{International Cosmic Ray
  Conference}, vol.~7, p. 256.
\newblock \doi{10.7529/ICRC2011/V07/1302}

\bibitem{2013ICRC...33.2768Z}
B.~{Zitzer}, {VERITAS Collaboration}, in \emph{International Cosmic Ray
  Conference}, \emph{International Cosmic Ray Conference}, vol.~33 (2013),
  \emph{International Cosmic Ray Conference}, vol.~33, p. 2768

\bibitem{2017ApJS..232....9M}
{MAGIC Collaboration}, M.L. {Ahnen}, S.~{Ansoldi}, et al., \apjs \textbf{232}(1), 9 (2017).
\newblock \doi{10.3847/1538-4365/aa8404}

\bibitem{2003PhRvL..90u1601M}
R.C. {Myers}, M.~{Pospelov}, Physical Review Letters \textbf{90}(21), 211601
  (2003).
\newblock \doi{10.1103/PhysRevLett.90.211601}

\bibitem{2003Natur.423..415C}
W.~{Coburn}, S.E. {Boggs}, \nat \textbf{423}, 415 (2003).
\newblock \doi{10.1038/nature01612}

\bibitem{2004MNRAS.350.1288R}
R.E. {Rutledge}, D.B. {Fox}, \mnras \textbf{350}, 1288 (2004).
\newblock \doi{10.1111/j.1365-2966.2004.07665.x}

\bibitem{2004ApJ...613.1088W}
C.~{Wigger}, W.~{Hajdas}, K.~{Arzner}, M.~{G{\"u}del}, A.~{Zehnder}, \apj
  \textbf{613}, 1088 (2004).
\newblock \doi{10.1086/423163}

\bibitem{2008PhRvD..78j3003M}
L.~{Maccione}, S.~{Liberati}, A.~{Celotti}, J.G. {Kirk}, P.~{Ubertini}, \prd
  \textbf{78}(10), 103003 (2008).
\newblock \doi{10.1103/PhysRevD.78.103003}

\bibitem{2007ApJS..169...75K}
E.~{Kalemci}, S.E. {Boggs}, C.~{Kouveliotou}, M.~{Finger}, M.G. {Baring}, \apjs
  \textbf{169}, 75 (2007).
\newblock \doi{10.1086/510676}

\bibitem{2007A&A...466..895M}
S.~{McGlynn}, D.J. {Clark}, A.J. {Dean}, L.~{Hanlon}, S.~{McBreen}, D.R.
  {Willis}, B.~{McBreen}, A.J. {Bird}, S.~{Foley}, \aap \textbf{466}, 895
  (2007).
\newblock \doi{10.1051/0004-6361:20066179}

\bibitem{2009ApJ...695L.208G}
D.~{G{\"o}tz}, P.~{Laurent}, F.~{Lebrun}, F.~{Daigne}, {\v Z}.~{Bo{\v s}njak},
  \apjl \textbf{695}, L208 (2009).
\newblock \doi{10.1088/0004-637X/695/2/L208}

\bibitem{2011ApJ...743L..30Y}
D.~{Yonetoku}, T.~{Murakami}, S.~{Gunji}, T.~{Mihara}, K.~{Toma},
  T.~{Sakashita}, Y.~{Morihara}, T.~{Takahashi}, N.~{Toukairin}, H.~{Fujimoto},
  Y.~{Kodama}, S.~{Kubo}, {IKAROS Demonstration Team}, \apjl \textbf{743}, L30
  (2011).
\newblock \doi{10.1088/2041-8205/743/2/L30}

\bibitem{2012ApJ...758L...1Y}
D.~{Yonetoku}, T.~{Murakami}, S.~{Gunji}, T.~{Mihara}, K.~{Toma},
  Y.~{Morihara}, T.~{Takahashi}, Y.~{Wakashima}, H.~{Yonemochi},
  T.~{Sakashita}, N.~{Toukairin}, H.~{Fujimoto}, Y.~{Kodama}, \apjl
  \textbf{758}, L1 (2012).
\newblock \doi{10.1088/2041-8205/758/1/L1}

\bibitem{2016ApJ...824L...9V}
H.K. {Vedantham}, V.~{Ravi}, K.~{Mooley}, D.~{Frail}, G.~{Hallinan}, S.R.
  {Kulkarni}, \apjl \textbf{824}(1), L9 (2016).
\newblock \doi{10.3847/2041-8205/824/1/L9}

\bibitem{2016ApJ...821L..22W}
P.K.G. {Williams}, E.~{Berger}, \apjl \textbf{821}(2), L22 (2016).
\newblock \doi{10.3847/2041-8205/821/2/L22}

\bibitem{2017Natur.541...58C}
S.~{Chatterjee}, C.J. {Law}, R.S. {Wharton}, S.~{Burke-Spolaor}, J.W.T.
  {Hessels}, G.C. {Bower}, J.M. {Cordes}, S.P. {Tendulkar}, C.G. {Bassa},
  P.~{Demorest}, B.J. {Butler}, A.~{Seymour}, P.~{Scholz}, M.W. {Abruzzo},
  S.~{Bogdanov}, V.M. {Kaspi}, A.~{Keimpema}, T.J.W. {Lazio}, B.~{Marcote},
  M.A. {McLaughlin}, Z.~{Paragi}, S.M. {Ransom}, M.~{Rupen}, L.G. {Spitler},
  H.J. {van Langevelde}, \nat \textbf{541}(7635), 58 (2017).
\newblock \doi{10.1038/nature20797}

\bibitem{2019ApJ...876L..23H}
J.W.T. {Hessels}, L.G. {Spitler}, A.D. {Seymour}, J.M. {Cordes}, D.~{Michilli},
  R.S. {Lynch}, K.~{Gourdji}, A.M. {Archibald}, C.G. {Bassa}, G.C. {Bower},
  S.~{Chatterjee}, L.~{Connor}, F.~{Crawford}, J.S. {Deneva}, V.~{Gajjar}, V.M.
  {Kaspi}, A.~{Keimpema}, C.J. {Law}, B.~{Marcote}, M.A. {McLaughlin},
  Z.~{Paragi}, E.~{Petroff}, S.M. {Ransom}, P.~{Scholz}, B.W. {Stappers}, S.P.
  {Tendulkar}, \apjl \textbf{876}(2), L23 (2019).
\newblock \doi{10.3847/2041-8213/ab13ae}

\bibitem{2017AdSpR..59..736B}
M.J. {Bentum}, L.~{Bonetti}, A.r.D.A.M. {Spallicci}, Advances in Space Research
  \textbf{59}(2), 736 (2017).
\newblock \doi{10.1016/j.asr.2016.10.018}

\bibitem{2014ApJ...783L..35D}
W.~{Deng}, B.~{Zhang}, \apjl \textbf{783}(2), L35 (2014).
\newblock \doi{10.1088/2041-8205/783/2/L35}

\bibitem{2016Natur.531..202S}
L.G. {Spitler}, P.~{Scholz}, J.W.T. {Hessels}, S.~{Bogdanov}, A.~{Brazier},
  F.~{Camilo}, S.~{Chatterjee}, J.M. {Cordes}, F.~{Crawford}, J.~{Deneva}, R.D.
  {Ferdman}, P.C.C. {Freire}, V.M. {Kaspi}, P.~{Lazarus}, R.~{Lynch}, E.C.
  {Madsen}, M.A. {McLaughlin}, C.~{Patel}, S.M. {Ransom}, A.~{Seymour}, I.H.
  {Stairs}, B.W. {Stappers}, J.~{van Leeuwen}, W.W. {Zhu}, \nat
  \textbf{531}(7593), 202 (2016).
\newblock \doi{10.1038/nature17168}

\bibitem{2017ApJ...834L...7T}
S.P. {Tendulkar}, C.G. {Bassa}, J.M. {Cordes}, G.C. {Bower}, C.J. {Law},
  S.~{Chatterjee}, E.A.K. {Adams}, S.~{Bogdanov}, S.~{Burke-Spolaor}, B.J.
  {Butler}, P.~{Demorest}, J.W.T. {Hessels}, V.M. {Kaspi}, T.J.W. {Lazio},
  N.~{Maddox}, B.~{Marcote}, M.A. {McLaughlin}, Z.~{Paragi}, S.M. {Ransom},
  P.~{Scholz}, A.~{Seymour}, L.G. {Spitler}, H.J. {van Langevelde}, R.S.
  {Wharton}, \apjl \textbf{834}(2), L7 (2017).
\newblock \doi{10.3847/2041-8213/834/2/L7}

\bibitem{2020Natur.577..190M}
B.~{Marcote}, K.~{Nimmo}, J.W.T. {Hessels}, et al., \nat \textbf{577}(7789), 190
  (2020).
\newblock \doi{10.1038/s41586-019-1866-z}

\bibitem{2019Sci...365..565B}
K.W. {Bannister}, A.T. {Deller}, C.~{Phillips}, et al., Science
  \textbf{365}(6453), 565 (2019).
\newblock \doi{10.1126/science.aaw5903}

\bibitem{2019Sci...366..231P}
J.X. {Prochaska}, J.P. {Macquart}, M.~{McQuinn}, S.~{Simha}, R.M. {Shannon},
  C.K. {Day}, L.~{Marnoch}, S.~{Ryder}, A.~{Deller}, K.W. {Bannister},
  S.~{Bhandari}, R.~{Bordoloi}, J.~{Bunton}, H.~{Cho}, C.~{Flynn}, E.K.
  {Mahony}, C.~{Phillips}, H.~{Qiu}, N.~{Tejos}, Science \textbf{366}(6462),
  231 (2019).
\newblock \doi{10.1126/science.aay0073}

\bibitem{2019Natur.572..352R}
V.~{Ravi}, M.~{Catha}, L.~{D'Addario}, S.G. {Djorgovski}, G.~{Hallinan},
  R.~{Hobbs}, J.~{Kocz}, S.R. {Kulkarni}, J.~{Shi}, H.K. {Vedantham},
  S.~{Weinreb}, D.P. {Woody}, \nat \textbf{572}(7769), 352 (2019).
\newblock \doi{10.1038/s41586-019-1389-7}

\bibitem{2020Natur.581..391M}
J.P. {Macquart}, J.X. {Prochaska}, M.~{McQuinn}, K.W. {Bannister},
  S.~{Bhandari}, C.K. {Day}, A.T. {Deller}, R.D. {Ekers}, C.W. {James},
  L.~{Marnoch}, S.~{Os{\l}owski}, C.~{Phillips}, S.D. {Ryder}, D.R. {Scott},
  R.M. {Shannon}, N.~{Tejos}, \nat \textbf{581}(7809), 391 (2020).
\newblock \doi{10.1038/s41586-020-2300-2}

\bibitem{2019MNRAS.485..648P}
J.X. {Prochaska}, Y.~{Zheng}, \mnras \textbf{485}(1), 648 (2019).
\newblock \doi{10.1093/mnras/stz261}

\bibitem{2015RAA....15.1629X}
J.~{Xu}, J.L. {Han}, Research in Astronomy and Astrophysics \textbf{15}(10),
  1629 (2015).
\newblock \doi{10.1088/1674-4527/15/10/002}

\bibitem{2018MNRAS.481.2320L}
R.~{Luo}, K.~{Lee}, D.R. {Lorimer}, B.~{Zhang}, \mnras \textbf{481}(2), 2320
  (2018).
\newblock \doi{10.1093/mnras/sty2364}

\bibitem{2006ApJ...651..142H}
A.M. {Hopkins}, J.F. {Beacom}, \apj \textbf{651}(1), 142 (2006).
\newblock \doi{10.1086/506610}

\bibitem{2008MNRAS.388.1487L}
L.X. {Li}, \mnras \textbf{388}(4), 1487 (2008).
\newblock \doi{10.1111/j.1365-2966.2008.13488.x}

\bibitem{2020A&A...641A...6P}
{Planck Collaboration}, N.~{Aghanim}, Y.~{Akrami}, et al., \aap \textbf{641}, A6
  (2020).
\newblock \doi{10.1051/0004-6361/201833910}

\bibitem{1998ApJ...503..518F}
M.~{Fukugita}, C.J. {Hogan}, P.J.E. {Peebles}, \apj \textbf{503}(2), 518
  (1998).
\newblock \doi{10.1086/306025}

\bibitem{2014Natur.513...71T}
R.B. {Tully}, H.~{Courtois}, Y.~{Hoffman}, D.~{Pomar{\`e}de}, Nature
  \textbf{513}, 71 (2014).
\newblock \doi{10.1038/nature13674}

\bibitem{2019PhRvD.100j4047M}
O.~{Minazzoli}, N.K. {Johnson-McDaniel}, M.~{Sakellariadou}, \prd
  \textbf{100}(10), 104047 (2019).
\newblock \doi{10.1103/PhysRevD.100.104047}

\bibitem{2016PhRvL.116o1101W}
Z.Y. {Wang}, R.Y. {Liu}, X.Y. {Wang}, Phys. Rev. Lett. \textbf{116}(15), 151101
  (2016).
\newblock \doi{10.1103/PhysRevLett.116.151101}

\bibitem{2019EPJC...79..185B}
S.~{Boran}, S.~{Desai}, E.O. {Kahya}, European Physical Journal C
  \textbf{79}(3), 185 (2019).
\newblock \doi{10.1140/epjc/s10052-019-6695-6}

\bibitem{2019PhRvD.100j3002L}
R.~{Laha}, \prd \textbf{100}(10), 103002 (2019).
\newblock \doi{10.1103/PhysRevD.100.103002}

\bibitem{2019JHEAp..22....1W}
J.J. {Wei}, B.B. {Zhang}, L.~{Shao}, H.~{Gao}, Y.~{Li}, Q.Q. {Yin}, X.F. {Wu},
  X.Y. {Wang}, B.~{Zhang}, Z.G. {Dai}, Journal of High Energy Astrophysics
  \textbf{22}, 1 (2019).
\newblock \doi{10.1016/j.jheap.2019.01.002}

\bibitem{2016JCAP...08..031W}
J.J. {Wei}, X.F. {Wu}, H.~{Gao}, P.~{M{\'e}sz{\'a}ros}, JCAP \textbf{8}, 031
  (2016).
\newblock \doi{10.1088/1475-7516/2016/08/031}

\bibitem{2016ApJ...827...75L}
X.~{Li}, Y.M. {Hu}, Y.Z. {Fan}, D.M. {Wei}, Astrophys. J. \textbf{827}, 75
  (2016).
\newblock \doi{10.3847/0004-637X/827/1/75}

\bibitem{2017ApJ...848L..13A}
B.P. {Abbott}, R.~{Abbott}, T.D. {Abbott}, F.~{Acernese}, K.~{Ackley},
  C.~{Adams}, T.~{Adams}, P.~{Addesso}, R.X. {Adhikari}, V.B. {Adya}, et~al.,
  Astrophys. J. \textbf{848}, L13 (2017).
\newblock \doi{10.3847/2041-8213/aa920c}

\bibitem{2017PhLB..770....8L}
M.~{Liu}, Z.~{Zhao}, X.~{You}, J.~{Lu}, L.~{Xu}, Phys. Lett. B \textbf{770}, 8
  (2017).
\newblock \doi{10.1016/j.physletb.2017.04.033}

\bibitem{2017ApJ...851L..18W}
H.~{Wang}, F.W. {Zhang}, Y.Z. {Wang}, Z.Q. {Shen}, Y.F. {Liang}, X.~{Li}, N.H.
  {Liao}, Z.P. {Jin}, Q.~{Yuan}, Y.C. {Zou}, Y.Z. {Fan}, D.M. {Wei}, Astrophys.
  J. \textbf{851}, L18 (2017).
\newblock \doi{10.3847/2041-8213/aa9e08}

\bibitem{2017JCAP...11..035W}
J.J. {Wei}, B.B. {Zhang}, X.F. {Wu}, H.~{Gao}, P.~{M{\'e}sz{\'a}ros},
  B.~{Zhang}, Z.G. {Dai}, S.N. {Zhang}, Z.H. {Zhu}, JCAP \textbf{11}, 035
  (2017).
\newblock \doi{10.1088/1475-7516/2017/11/035}

\bibitem{2018PhRvD..97h3013S}
I.M. {Shoemaker}, K.~{Murase}, \prd \textbf{97}(8), 083013 (2018).
\newblock \doi{10.1103/PhysRevD.97.083013}

\bibitem{2018PhRvD..97d1501B}
S.~{Boran}, S.~{Desai}, E.O. {Kahya}, R.P. {Woodard}, \prd \textbf{97}(4),
  041501 (2018).
\newblock \doi{10.1103/PhysRevD.97.041501}

\bibitem{2020ApJ...900...31Y}
L.~{Yao}, Z.~{Zhao}, Y.~{Han}, J.~{Wang}, T.~{Liu}, M.~{Liu}, \apj
  \textbf{900}(1), 31 (2020).
\newblock \doi{10.3847/1538-4357/abab02}

\bibitem{1999BASI...27..627S}
C.~{Sivaram}, Bulletin of the Astronomical Society of India \textbf{27}, 627
  (1999)

\bibitem{2016MNRAS.460.2282S}
Y.~{Sang}, H.N. {Lin}, Z.~{Chang}, Mon. Not. R. Astron. Soc. \textbf{460}, 2282
  (2016).
\newblock \doi{10.1093/mnras/stw1136}

\bibitem{2016JHEAp...9...35L}
Z.X. {Luo}, B.~{Zhang}, J.J. {Wei}, X.F. {Wu}, JHEAp \textbf{9}, 35 (2016).
\newblock \doi{10.1016/j.jheap.2016.04.001}

\bibitem{2018ApJ...860..173Y}
H.~{Yu}, S.Q. {Xi}, F.Y. {Wang}, Astrophys. J. \textbf{860}, 173 (2018).
\newblock \doi{10.3847/1538-4357/aac2e3}

\bibitem{2016ApJ...820L..31T}
S.J. {Tingay}, D.L. {Kaplan}, Astrophys. J. \textbf{820}, L31 (2016).
\newblock \doi{10.3847/2041-8205/820/2/L31}

\bibitem{2016ApJ...821L...2N}
A.~{Nusser}, Astrophys. J. \textbf{821}, L2 (2016).
\newblock \doi{10.3847/2041-8205/821/1/L2}

\bibitem{2020PDU....2900571W}
D.~{Wang}, Z.~{Li}, J.~{Zhang}, Physics of the Dark Universe \textbf{29},
  100571 (2020).
\newblock \doi{10.1016/j.dark.2020.100571}

\bibitem{2016ApJ...818L...2W}
J.J. {Wei}, J.S. {Wang}, H.~{Gao}, X.F. {Wu}, Astrophys. J. \textbf{818}, L2
  (2016).
\newblock \doi{10.3847/2041-8205/818/1/L2}

\bibitem{2016PhRvD..94j1501Y}
Y.P. {Yang}, B.~{Zhang}, Phys. Rev. D \textbf{94}(10), 101501 (2016).
\newblock \doi{10.1103/PhysRevD.94.101501}

\bibitem{2017ApJ...837..134Z}
Y.~{Zhang}, B.~{Gong}, Astrophys. J. \textbf{837}, 134 (2017).
\newblock \doi{10.3847/1538-4357/aa61fb}

\bibitem{2018EPJC...78...86D}
S.~{Desai}, E.~{Kahya}, European Physical Journal C \textbf{78}, 86 (2018).
\newblock \doi{10.1140/epjc/s10052-018-5571-0}

\bibitem{2018ApJ...861...66L}
C.~{Leung}, B.~{Hu}, S.~{Harris}, A.~{Brown}, J.~{Gallicchio}, H.~{Nguyen},
  Astrophys. J. \textbf{861}, 66 (2018).
\newblock \doi{10.3847/1538-4357/aac954}

\bibitem{2016PhLB..756..265K}
E.O. {Kahya}, S.~{Desai}, Phys. Lett. B \textbf{756}, 265 (2016).
\newblock \doi{10.1016/j.physletb.2016.03.033}

\bibitem{2020MNRAS.499L..53Y}
S.C. {Yang}, W.B. {Han}, G.~{Wang}, \mnras \textbf{499}(1), L53 (2020).
\newblock \doi{10.1093/mnrasl/slaa143}

\bibitem{2018EPJC...78..692Y}
H.~{Yu}, F.Y. {Wang}, European Physical Journal C \textbf{78}(9), 692 (2018).
\newblock \doi{10.1140/epjc/s10052-018-6162-9}

\bibitem{2019arXiv191206891M}
O.~{Minazzoli}, arXiv e-prints arXiv:1912.06891 (2019)

\bibitem{2017PhRvD..95j3004W}
X.F. {Wu}, J.J. {Wei}, M.X. {Lan}, H.~{Gao}, Z.G. {Dai}, P.~{M{\'e}sz{\'a}ros},
  Phys. Rev. D \textbf{95}(10), 103004 (2017).
\newblock \doi{10.1103/PhysRevD.95.103004}

\bibitem{2020EPJP..135..527W}
J.J. {Wei}, X.F. {Wu}, European Physical Journal Plus \textbf{135}(6), 527
  (2020).
\newblock \doi{10.1140/epjp/s13360-020-00554-x}

\bibitem{2020MNRAS.493.1782Y}
S.X. {Yi}, Y.C. {Zou}, X.~{Yang}, B.~{Liao}, S.W. {Wei}, \mnras
  \textbf{493}(2), 1782 (2020).
\newblock \doi{10.1093/mnras/staa369}

\bibitem{2020MNRAS.498.4295Y}
S.X. {Yi}, Y.C. {Zou}, J.J. {Wei}, Q.Q. {Zhou}, \mnras \textbf{498}(3), 4295
  (2020).
\newblock \doi{10.1093/mnras/staa2686}

\bibitem{2019Natur.575..464A}
H.~{Abdalla}, R.~{Adam}, F.~{Aharonian}, et al., \nat \textbf{575}(7783), 464 (2019).
\newblock \doi{10.1038/s41586-019-1743-9}

\bibitem{2019Natur.575..448Z}
B.~{Zhang}, \nat \textbf{575}(7783), 448 (2019).
\newblock \doi{10.1038/d41586-019-03503-6}

\bibitem{2017NewAR..76....1M}
M.L. {McConnell}, \nar \textbf{76}, 1 (2017).
\newblock \doi{10.1016/j.newar.2016.11.001}

\bibitem{2000CQGra..17.4999G}
S.~{Gao}, R.M. {Wald}, Classical and Quantum Gravity \textbf{17}(24), 4999
  (2000).
\newblock \doi{10.1088/0264-9381/17/24/305}

\bibitem{2017NatAs...1E..36H}
Y.~{Hoffman}, D.~{Pomar{\`e}de}, R.B. {Tully}, H.M. {Courtois}, Nature
  Astronomy \textbf{1}, 0036 (2017).
\newblock \doi{10.1038/s41550-016-0036}

\end{thebibliography}

\raggedend
%



\end{document}